\pdfoutput=1
\documentclass[a4paper,11pt]{article}
\usepackage{jinstpub} %
\pdfmapline{+newtxtte newtxtt " txttECEncodingE ReEncodeFont " <[txttEec.enc <newtxtt.pfb}
\pdfmapline{+newtxbtte newtxbtt " txttECEncodingE ReEncodeFont " <[txttEec.enc <newtxbtt.pfb}
\pdfmapline{+newtxttsle newtxtt " .167 SlantFont txttECEncodingE ReEncodeFont " <[txttEec.enc <newtxtt.pfb}
\pdfmapline{+newtxbttsle newtxbtt " .167 SlantFont txttECEncodingE ReEncodeFont " <[txttEec.enc <newtxbtt.pfb}
\pdfmapline{+newtxttsce newtxttsc " txttECEncodingE ReEncodeFont " <[txttEec.enc <newtxttsc.pfb}
\pdfmapline{+newtxbttsce newtxbttsc " txttECEncodingE ReEncodeFont " <[txttEec.enc <newtxbttsc.pfb}
 \usepackage{xspace}
\newcommand{\pythia}{\mbox{\textsc{Pythia}}\xspace}
\newcommand{\evtgen}{\mbox{\textsc{EvtGen}}\xspace}
\newcommand{\photos}{\mbox{\textsc{Photos}}\xspace}
\newcommand{\geant}{\mbox{\textsc{Geant4}}\xspace}

\title{Real-time lepton identification at LHCb in Run 3 using Lipschitz neural networks}

\IfFileExists{latexml.sty}{\usepackage{latexml}}{\newif\iflatexml\latexmlfalse}
\iflatexml
\author{Adrian Casais Vidal\\Massachusetts Institute of Technology, Cambridge, MA, USA\\Corresponding author: acasaisv@mit.edu
\and Kate A. Richardson\\Massachusetts Institute of Technology, Cambridge, MA, USA\\NSF AI Institute for Artificial Intelligence and Fundamental Interactions
\and Maarten Van Veghel\\Universiteit Maastricht, Maastricht, Netherlands\\Nikhef National Institute for Subatomic Physics, Amsterdam, Netherlands
\and Marco Santimaria\\INFN Laboratori Nazionali di Frascati, Frascati, Italy
\and Marianna Fontana\\INFN Sezione di Bologna, Bologna, Italy
\and Mike Williams\\Massachusetts Institute of Technology, Cambridge, MA, USA\\NSF AI Institute for Artificial Intelligence and Fundamental Interactions}
\else
\author[a,1]{Adrian Casais Vidal\note{Corresponding author.}}
\author[a,b]{Kate A. Richardson}
\author[c,d]{Maarten Van Veghel}
\author[e]{Marco Santimaria}
\author[f]{Marianna Fontana}
\author[a,b]{Mike Williams}
\affiliation[a]{Massachusetts Institute of Technology, Cambridge, MA, USA}
\affiliation[b]{NSF AI Institute for Artificial Intelligence and Fundamental Interactions}
\affiliation[c]{Universiteit Maastricht, Maastricht, Netherlands}
\affiliation[d]{Nikhef National Institute for Subatomic Physics, Amsterdam, Netherlands}
\affiliation[e]{INFN Laboratori Nazionali di Frascati, Frascati, Italy}
\affiliation[f]{INFN Sezione di Bologna, Bologna, Italy}
\fi

\iflatexml\else
\emailAdd{acasaisv@mit.edu}
\fi

\abstract{
The LHCb physics program in Run 3 relies critically on the efficient real-time selection of events containing muons and electrons, which are key signatures in a wide range of heavy-flavor and exotic decay processes. The LHCb Run 3 detector now operates with a fully software-based trigger that processes the complete detector readout at the LHC bunch-crossing rate, with the first trigger stage executed on GPUs. In this environment, particle-identification algorithms must achieve high efficiency and background rejection while satisfying stringent constraints on throughput and memory footprint. We present algorithms for muon and electron identification in the LHCb Run 3 GPU trigger based on Lipschitz-constrained neural networks. Separate networks are developed for muons and electrons and are trained using simulated events. Their performance is evaluated relative to the previous baseline algorithms, demonstrating improved discrimination across a wide range of kinematic regions while remaining compatible with the requirements of real-time GPU execution.} %

\begin{document}
\maketitle
\flushbottom

\section{Introduction}
\label{sec:intro}

The LHCb detector~\cite{lhcb,LHCb:2014set} located at the Large Hadron Collider (LHC) at CERN is a single-arm forward spectrometer covering the pseudorapidity range ${2 < \eta < 5}$.\footnote{Natural units, in which $c = 1$, are used throughout this article.}
Many of the key measurements in the LHCb physics program involve final states containing leptons, in particular muons and electrons, including a broad range of heavy-flavor and exotic decay processes. 
Therefore, the ability to efficiently identify leptons at the trigger level while simultaneously minimizing the misidentified hadronic background rate is an essential element of detector performance.

During Runs 1 and 2 (2010--2018), LHCb collected approximately 9~fb$^{-1}$ of integrated luminosity and delivered a broad physics program. 
The LHCb Run 3 detector, installed during LHC Long Shutdown 2 (2019--2022), was designed to substantially extend this reach by enabling operation at a much higher instantaneous luminosity and by reading out the full detector at the 40~MHz LHC bunch-crossing rate~\cite{LHCbUpgradeI}. 
In Run 3 and Run 4, the upgraded experiment is expected to accumulate at least 50~fb$^{-1}$ of data.
A defining feature of the Run 3 detector is the removal of the hardware trigger and the adoption of a triggerless readout architecture~\cite{LHCb:2014dlo}. 
All subdetectors are now read out for every bunch crossing, resulting in an input rate of approximately 30~MHz of inelastic proton–proton collision events to the software trigger (referred to as the high-level trigger or HLT). 
Event selection in the HLT uses a two-stage approach: a first stage (HLT1) executed on a large GPU farm, followed by a second stage (HLT2) that performs full offline-quality event reconstruction on a CPU farm.
HLT1 trigger algorithms must process immense data volumes in real time, while remaining compatible with the memory constraints imposed by the GPU hardware. 

The upgraded detector provides the information required for lepton identification already at the first trigger stage. 
A new high-granularity pixel vertex detector (VELO) and an upgraded tracking system enable fast and efficient reconstruction of charged particles~\cite{LHCb:2014uqj}, while the calorimeter~\cite{ecal-performance} and muon~\cite{muon-performancerun12,muon-performancerun3} systems provide complementary information for electron and muon identification, respectively. 
The particle-identification sub-detectors were upgraded to sustain the full readout rate, ensuring that lepton-related observables are available for trigger-level decisions.

Trigger-level particle identification (PID) in Run 3 has traditionally used selections based on simple cuts on reconstructed features, chosen to be efficient and computationally inexpensive. 
While such approaches are robust, machine-learning methods offer improved discrimination power---but must be carefully designed to also be robust, in addition to satisfying the stringent throughput and memory footprint requirements of the real-time GPU trigger environment.
In this article, we present neural-network-based algorithms for muon and electron identification in the LHCb Run 3 trigger. 
The algorithms are based on Lipschitz-constrained neural networks, which explicitly bound the sensitivity of the network output to variations in the input features and enable efficient model implementations~\cite{Kitouni:2021fkh}. 
Separate networks are developed for muon and electron identification and are trained using simulated events. 
Although the algorithms have been deployed and validated in the Run 3 trigger system, the results presented here are based exclusively on simulated data.

\section{Real-time analysis and the Allen framework}

The first stage of the LHCb software trigger (HLT1) is executed on a large GPU farm  using the custom Allen framework~\cite{Allen}. 
HLT1 performs a fast partial reconstruction of every proton–proton collision that produces particles within the LHCb acceptance at the full input rate of approximately 30~MHz (corresponding to about 5~TB/s), with the goal of selecting potentially interesting events for further processing. 
Its primary tasks include the reconstruction of charged particle tracks, primary and secondary vertex finding, muon identification, and the application of inclusive trigger selections such as two-track topological triggers~\cite{Delaney:2023swp}. This stage reduces the event rate  down to $\mathcal{O}(1\,{\rm MHz})$. 

The use of GPUs for HLT1 represents a major architectural change for LHCb and enables the experiment to exploit massive parallelism for real-time event reconstruction. 
The Allen framework is implemented in CUDA C++ and follows a dataflow-based execution model in which reconstruction algorithms are organized into a sequence of interdependent GPU kernels. 
Each kernel performs a specific reconstruction task, such as hit decoding, track finding, or vertex fitting, and is executed concurrently over large numbers of events and detector elements.

A key design principle of Allen is the optimization of throughput and memory usage. 
The reconstruction sequence is easily configurable, and a scheduler resolves data dependencies between algorithms to ensure each kernel is executed only when its inputs are available and without redundant computation.
To avoid costly dynamic memory allocations during processing, Allen employs a custom memory management strategy in which large memory pools are pre-allocated on the GPU at initialization. 
Data structures are stored in a structure-of-arrays format to enable coalesced memory accesses and efficient use of GPU memory bandwidth.
These design choices are essential for processing the high event and track rates in real time as required in HLT1.
In addition to charged-particle reconstruction, HLT1 has been extended to include algorithms for neutral object reconstruction, such as photons and $\pi^0$ mesons, electron identification and bremsstrahlung recovery, and downstream tracking to recover tracks that do not traverse the vertex detector~\cite{Kholoimov:2025cqe}.

Machine-learning techniques are playing an increasingly important role in HLT1. 
The use of Lipschitz-constrained neural networks provides a well-controlled model behavior and enables compact implementations that are well suited to execution on GPUs~\cite{Kitouni:2021fkh,Delaney:2023swp}. 
In the context of HLT1, such models must process a huge rate of reconstructed tracks while maintaining a small memory footprint and minimal impact on overall throughput.
The neural networks for muon and electron identification presented in this work are implemented as GPU kernels within the Allen reconstruction sequence. 
Network inference is performed in a batched manner, evaluating large collections of tracks from multiple events simultaneously to maximize GPU occupancy and arithmetic intensity. 
This approach allows the evaluation of these neural networks on all tracks with associated muon and/or electron information with minimal impact on the HLT1 throughput or memory footprint. 
Finally, events accepted by HLT1 are buffered to disk and processed asynchronously by HLT2.
The work presented here focuses exclusively on the design and performance of the muon and electron identification algorithms deployed in HLT1.

\section{Lipschitz-constrained neural networks for lepton identification}

Muon and electron identification in the HLT1 trigger is performed using neural networks with explicit Lipschitz constraints~\cite{Kitouni:2021fkh}. 
A Lipschitz-constrained network bounds the sensitivity of its output to changes in its input: for a network $f$ with Lipschitz constant $K$, $\lVert f(x) - f(x')\rVert \le K \lVert x - x'\rVert_1$ for all inputs $x,x'$, such that no small perturbation of the input features can produce a disproportionately large change in the network output. This bounded sensitivity is advantageous in a real-time trigger environment, where detector conditions and event properties can vary.
In addition, Lipschitz-constrained architectures enable compact model designs that are well suited to execution on GPUs under stringent throughput and memory footprint constraints.

Separate neural networks are trained for muon and electron identification. 
In both cases, fully connected feed-forward architectures are employed, with the depth and width chosen to balance classification performance against the computational requirements of real-time inference in HLT1. 
To enforce the Lipschitz constraint, the weight-normalization scheme of Ref.~\cite{Kitouni:2021fkh} is applied to each layer. In the first layer, each weight is individually bounded, while in subsequent layers the sum of the absolute weights entering any output node is bounded. GroupSort activation functions~\cite{Anil:2018groupsort} are used throughout and do not increase the Lipschitz constant. The full network is therefore Lipschitz with a constant $K$ given by the product of the per-layer bounds.
The muon identification network consists of five fully connected layers with width 10. %
The electron identification network uses a shallower architecture with four layers of width 12. %
For both networks, depth and width were chosen to minimize the network size without degrading the validation performance within the HLT1 timing budget; for the electron network, deeper architectures bring no significant gain.
These architectures are sufficiently small to ensure fast evaluation for every reconstructed track in HLT1.
All muon inputs, and all electron inputs except the two energy-to-momentum ratios, have long tails and are therefore transformed to their logarithms. Each input is then rescaled linearly to the interval $[0,1]$ based on the bulk of its distribution: candidates in the far tails of any input are removed from the training samples, and the minimum and maximum of each input over the remaining signal and background candidates define the rescaling.
The same constants are used online, where inputs are not clipped. 
This preprocessing ensures that features contribute on comparable scales and facilitates stable training under Lipschitz constraints. 
For each architecture, the Lipschitz constant is chosen to be the smallest value that maintains optimal classification performance on validation data, providing robustness without over-constraining the network.

The neural networks are trained offline using simulated events, generated with \pythia~\cite{Sjostrand:2007gs}. Decays of unstable particles are described by \evtgen~\cite{Lange:2001uf}, in which final-state radiation is generated using \photos~\cite{Golonka:2005pn}. The interaction of the generated particles with the detector, and its response, are implemented using the \geant toolkit~\cite{Agostinelli:2002hh}.
For muon identification, the signal class consists of genuine muons originating from $J/\psi \to \mu^+\mu^-$ and $\Upsilon$(1S) $\to \mu^+\mu^-$ decays. 
The background class is composed primarily of protons, which serve as a representative sample of stable hadrons. 
Although pions dominate the hadronic background in LHCb data, their in-flight decays introduce a bimodal behavior that can complicate training. 
Using protons as the primary background encourages the network to learn features that distinguish muons from stable hadrons, which is found to generalize well to pion rejection. 
The trained models are validated against pion backgrounds to confirm robust performance.
For electron identification, signal electrons are taken from simulated $J/\psi \to e^+e^-$ and $\Upsilon$(1S) $\to e^+e^-$ decays. 
The background class consists primarily of charged pions and kaons, which can mimic electron signatures in the calorimeter. %

Both networks are trained by minimizing the binary cross-entropy between the network output, passed through a sigmoid function, and the true class. The Adam optimizer~\cite{Kingma:2014vow} is used with batches of 128 candidates and a learning rate that decreases exponentially at each step from $10^{-2}$ to $10^{-3}$ over 50 epochs. A quarter of the training sample is held out for validation. To account for the different sizes of the two classes, each background candidate is weighted in the loss by the fraction of signal candidates in its batch. The training samples contain about $1.6\times10^{5}$ electrons and $1.9\times10^{6}$ pions for the electron network, and about $8.9\times10^{4}$ muons and $2.2\times10^{4}$ protons for the muon network. The Lipschitz normalization is applied at every training step, so that the bound holds throughout training.

\section{Muon identification}

Muon identification at LHCb relies on information from the muon detector system \cite{muon-performancerun12,muon-performancerun3}, which consists of four stations (M2--M5) of multi-wire proportional chambers located downstream of the calorimeters and interleaved with iron absorbers. 
Muon candidates are identified by matching charged-particle tracks reconstructed in the tracking system to hits recorded in the muon stations \cite{muidrun12}. 
The principal challenge is to distinguish genuine muons from hadrons, primarily pions and kaons, that either penetrate the calorimeters and absorbers without showering, or that decay in flight, producing secondary muons.

A central input to the muon identification is the $\chi^2_{\rm Corr}$ matching operator~\cite{LHCbMuonID2020}, which quantifies the compatibility between the extrapolated track trajectory and the observed muon hits while accounting for the effects of multiple scattering in the detector material, which can make the true trajectory deviate from the naive idealized path. 
In addition, the variable $D^2$ is used, defined as the sum of squared residuals between the extrapolated track and the muon hits without accounting for the multiple-scattering deviations. 
This complementary observable provides additional discrimination power, particularly for high-momentum tracks where multiple scattering is less significant.

\begin{figure}
    \centering
\includegraphics[width=0.32\textwidth]{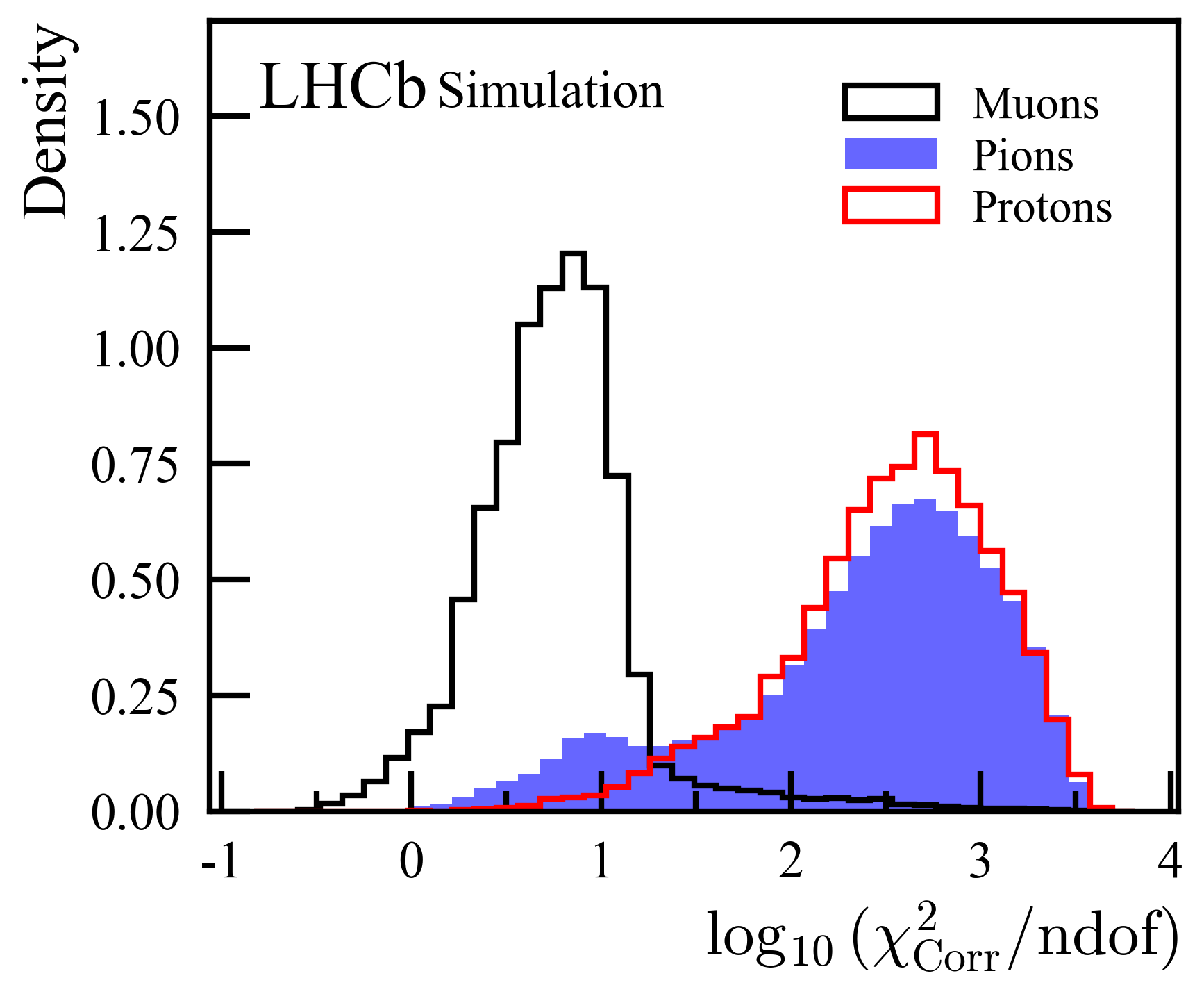}
\includegraphics[width=0.32\textwidth]{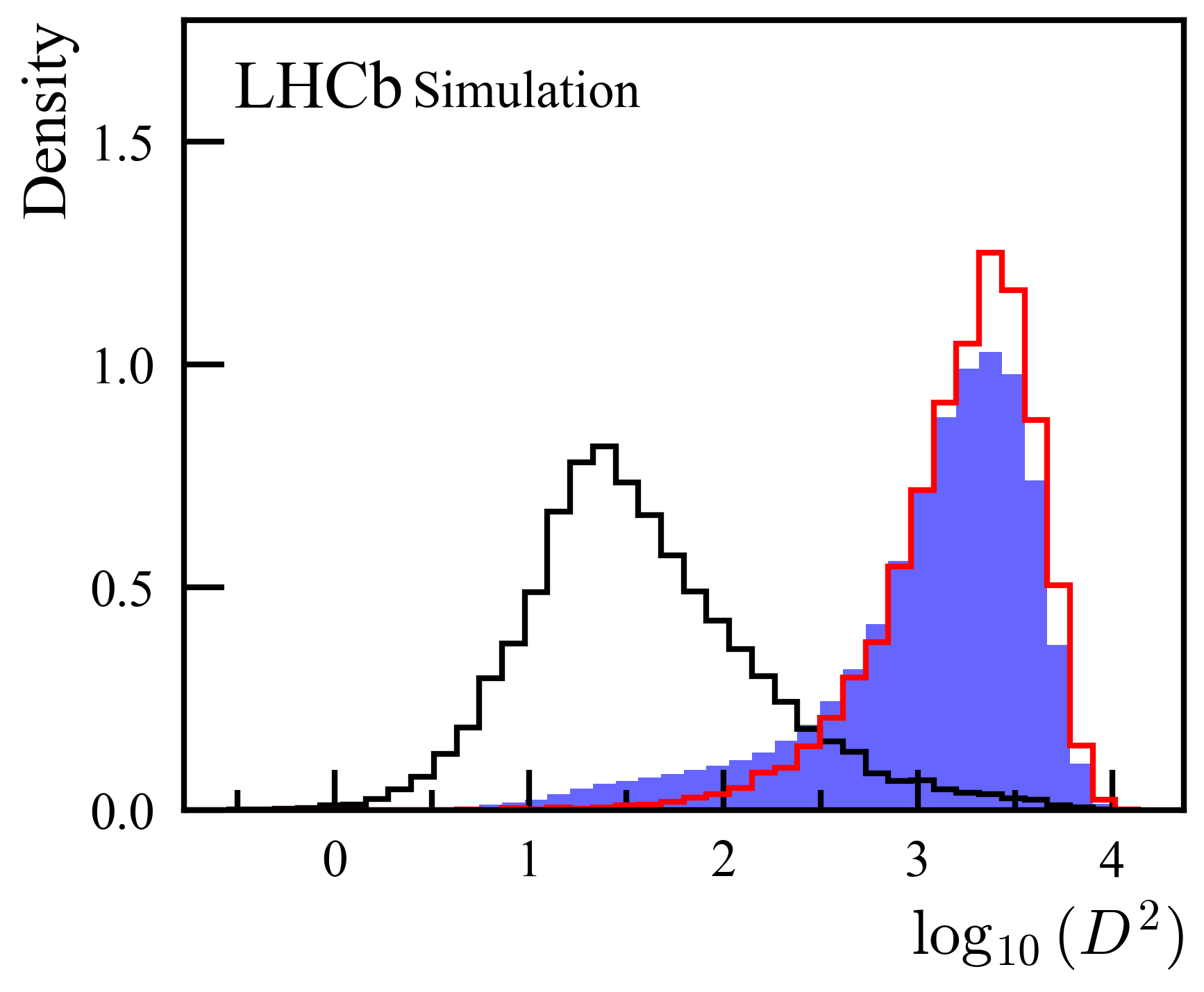}
\includegraphics[width=0.32\textwidth]{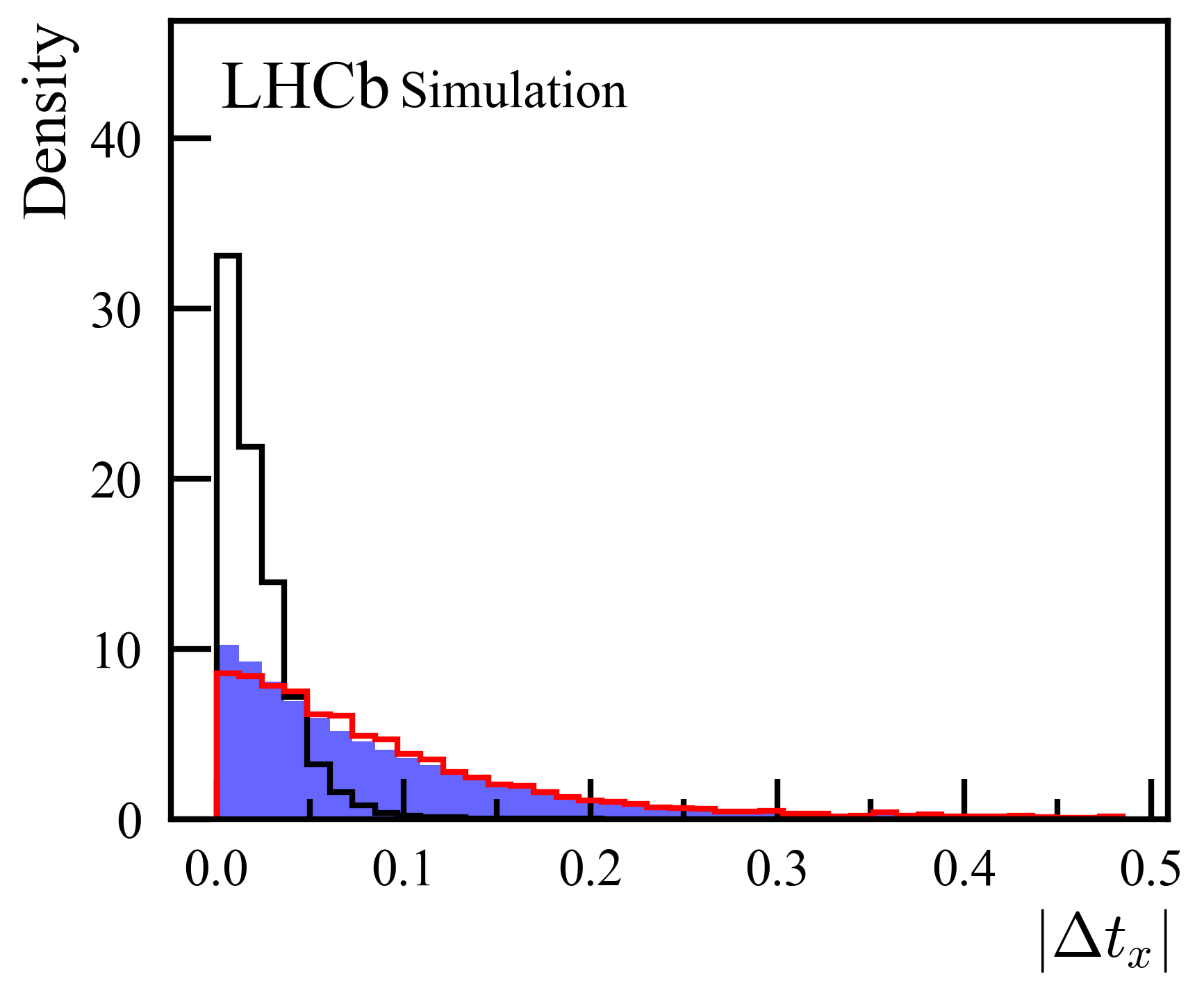}
\includegraphics[width=0.32\textwidth]{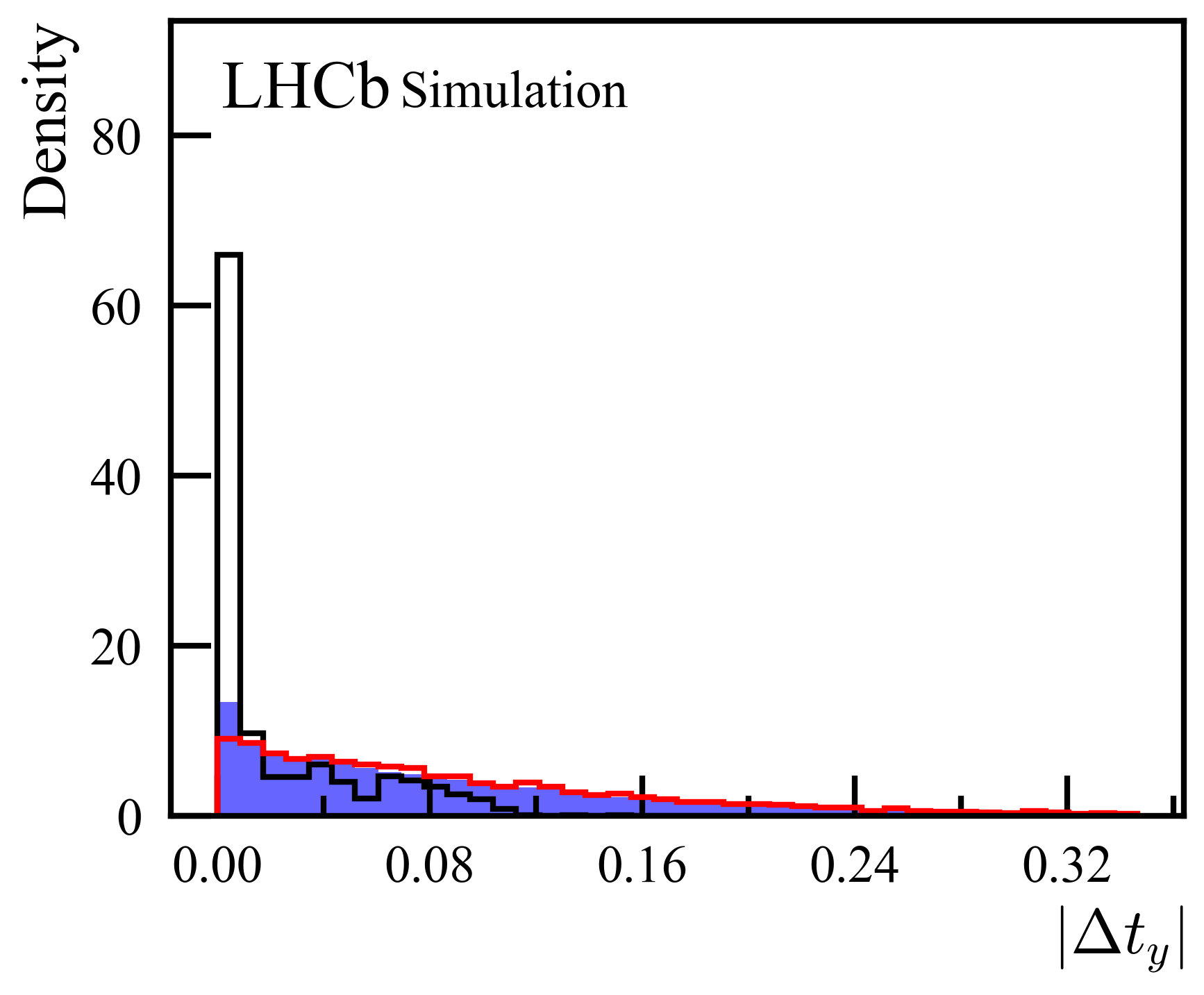}
\includegraphics[width=0.32\textwidth]{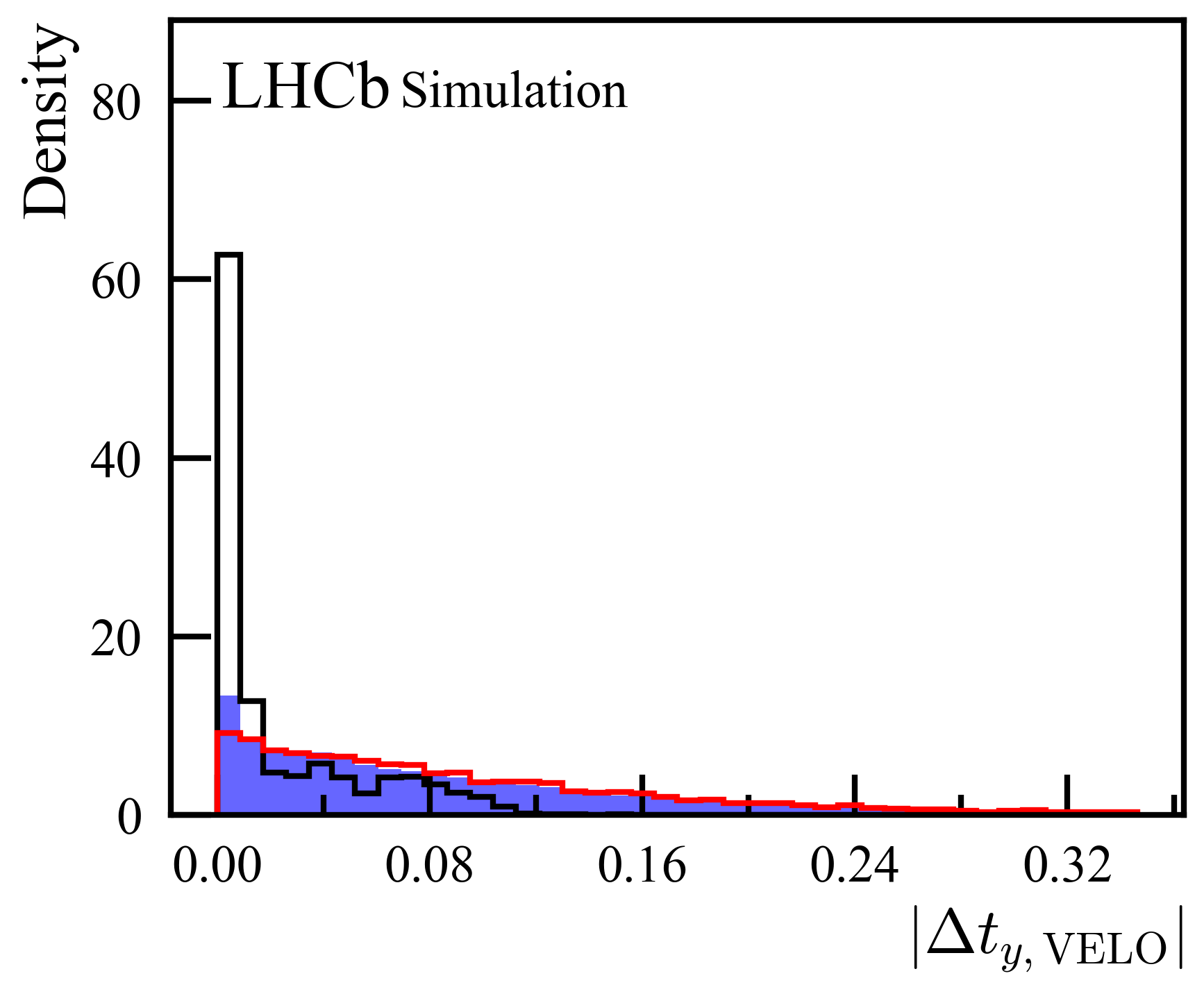}
\caption{\label{fig:muon-features} Distributions of the five input features used by the muon-identification Lipschitz neural network (prior to rescaling), shown for simulated muons (black), pions (blue), and protons (red) in the momentum bin $p\in[6,10]\,\mathrm{GeV}$.}
\end{figure}

Further discriminating information is obtained from the comparison of track slopes measured at different detector locations. 
Track slopes $t_x$ and $t_y$, defined as the derivatives $dx/dz$ and $dy/dz$ with $\hat{z}$ defining the beam direction, are evaluated at the Scintillating Fiber (SciFi) tracker, the VELO, and the muon system. From these, differences between slopes are constructed, including
\[
\Delta t_{x,y} = \left| t_{x,y}^{\mathrm{MUON}} - t_{x,y}^{\mathrm{SciFi}} \right|
\quad \text{and} \quad
\Delta t_{y,\mathrm{VELO}} = \left| t_{y}^{\mathrm{MUON}} - t_{y}^{\mathrm{VELO}} \right|.
\]
Differences in the $x$-component between the VELO and the muon system are not used, as the deflection induced by the dipole magnet would make this quantity primarily sensitive to the momentum rather than PID.
The physical motivation for these slope-difference variables is that no strong magnetic field is present between the SciFi tracker and the muon system, hence the track direction is expected to remain approximately constant for a genuine muon. 
Similarly, in the non-bending ($yz$) plane, the track slope should remain stable along the full trajectory from the VELO to the muon stations. 
In contrast, hadronic interactions or decays in flight introduce kinks and discontinuities in the reconstructed track, leading to larger slope differences.
These effects are particularly pronounced for pions, which %
frequently decay before reaching the muon system.
Figure~\ref{fig:muon-features} shows the distributions of the five input features used in the muon-identification Lipschitz neural network (prior to rescaling).

 \section{Electron identification}

Electron identification at LHCb relies primarily on information from the electromagnetic calorimeter (ECAL), a shashlik-type sampling calorimeter composed of alternating layers of lead absorber and scintillator tiles~\cite{ecal-performance}. 
Electrons deposit most of their energy in the ECAL through electromagnetic showering, while charged hadrons typically deposit only a small fraction of their energy. 
Therefore, the ECAL provides both energy measurements and spatial information that are essential for discriminating electrons from hadronic backgrounds.

In the HLT1 trigger, electron identification is performed by extrapolating tracks reconstructed in the SciFi tracker to the ECAL surface and associating them with nearby calorimeter energy deposits. 
For each matched track-cluster pair, a set of discriminating observables is computed. 
The barycenter of the ECAL cluster is defined as the energy-weighted centroid of the contributing cells,
\[
x_b = \frac{1}{\epsilon_T} \sum_i \epsilon_i x_i,
\qquad
y_b = \frac{1}{\epsilon_T} \sum_i \epsilon_i y_i,
\]
where $\epsilon_i$ is the energy deposited in cell $i$ at position $(x_i,y_i)$ and $\epsilon_T = \sum_i \epsilon_i$ is the total cluster energy. 
The squared distance between the extrapolated track position and the cluster barycenter, normalized to the ECAL cell width, $w$, 
\[
\Delta_b^2 = \left(\frac{x_{\mathrm{SciFi}} - x_b}{w} \right)^2 + \left(\frac{y_{\mathrm{SciFi}} - y_b}{w}\right)^2,
\]
provides discrimination against neutral particles, such as photons and neutral hadrons, which do not produce charged tracks.

One of the most powerful discriminating observables is the energy-over-momentum ratio, $E/p$, where $E$ is the ECAL energy associated with the track and $p$ is the track momentum. 
For electrons, $E/p$ is expected to be close to unity, while hadrons typically yield much smaller values. 
Both an $E/p$ value computed using only cells crossed by the extrapolated track and a variant using a $3\times3$ cell region around the barycenter are used, providing sensitivity to both energy containment and shower shape.

Additional information is obtained from the transverse shape of the ECAL energy deposit. Second moments of the energy distribution are computed as
\begin{align*}
\mathcal{S}_{xx} &= \frac{\sum_i \epsilon_i (x_i - x_b)^2}{\epsilon_T \, w^2}, \\
\mathcal{S}_{yy} &= \frac{\sum_i \epsilon_i (y_i - y_b)^2}{\epsilon_T \, w^2}, \\
\tilde{\mathcal{S}}_{xy} &= \frac{\sum_i \epsilon_i (x_i - x_b)(y_i - y_b)}{\epsilon_T}.
\end{align*}
These shower-shape variables characterize the transverse spread and symmetry of the energy deposition. 
Electrons typically produce compact, approximately symmetric electromagnetic showers, while hadronic showers tend to be broader and more irregular.

\begin{figure}
    \centering
\includegraphics[width=0.32\textwidth]{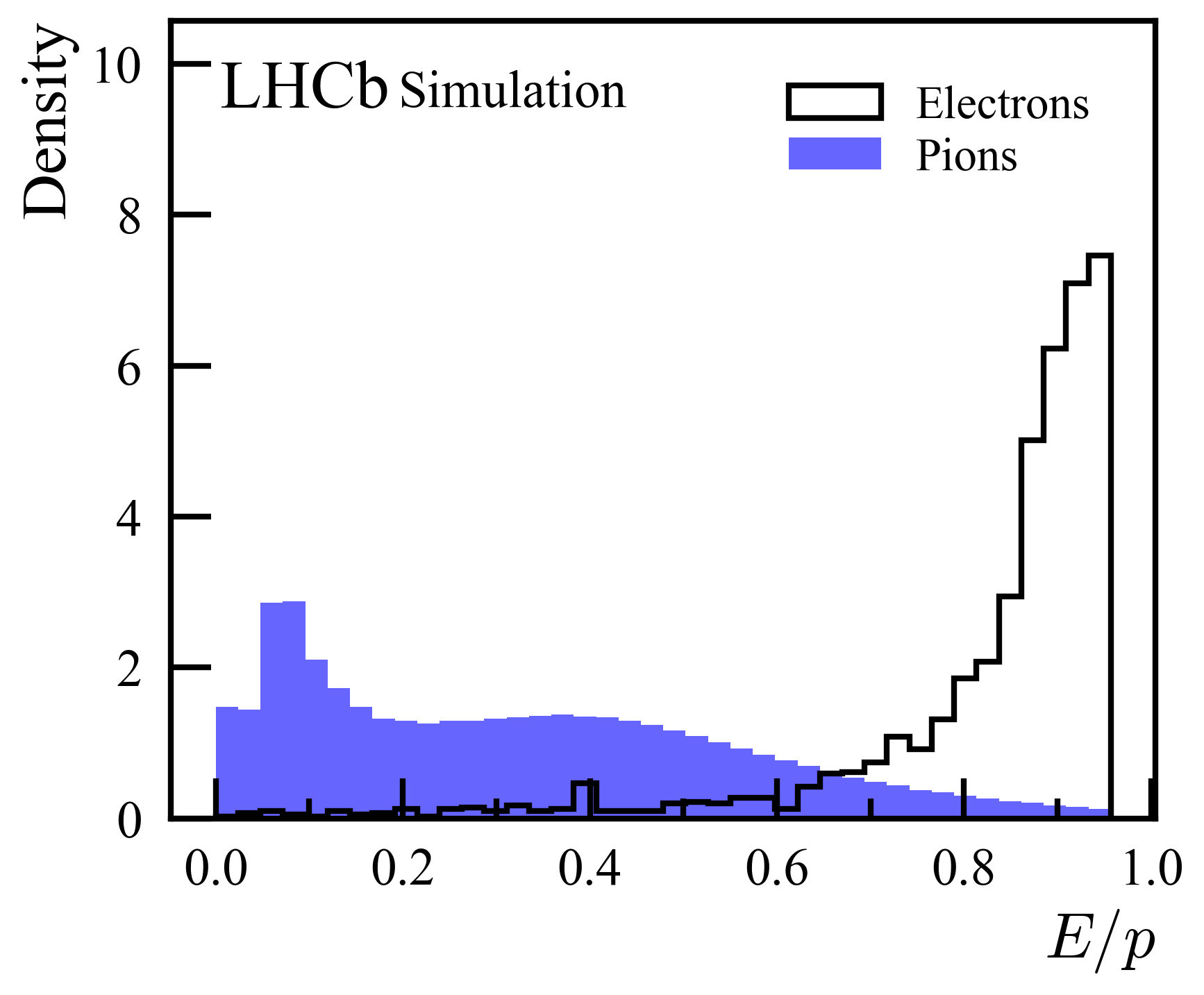}
\includegraphics[width=0.32\textwidth]{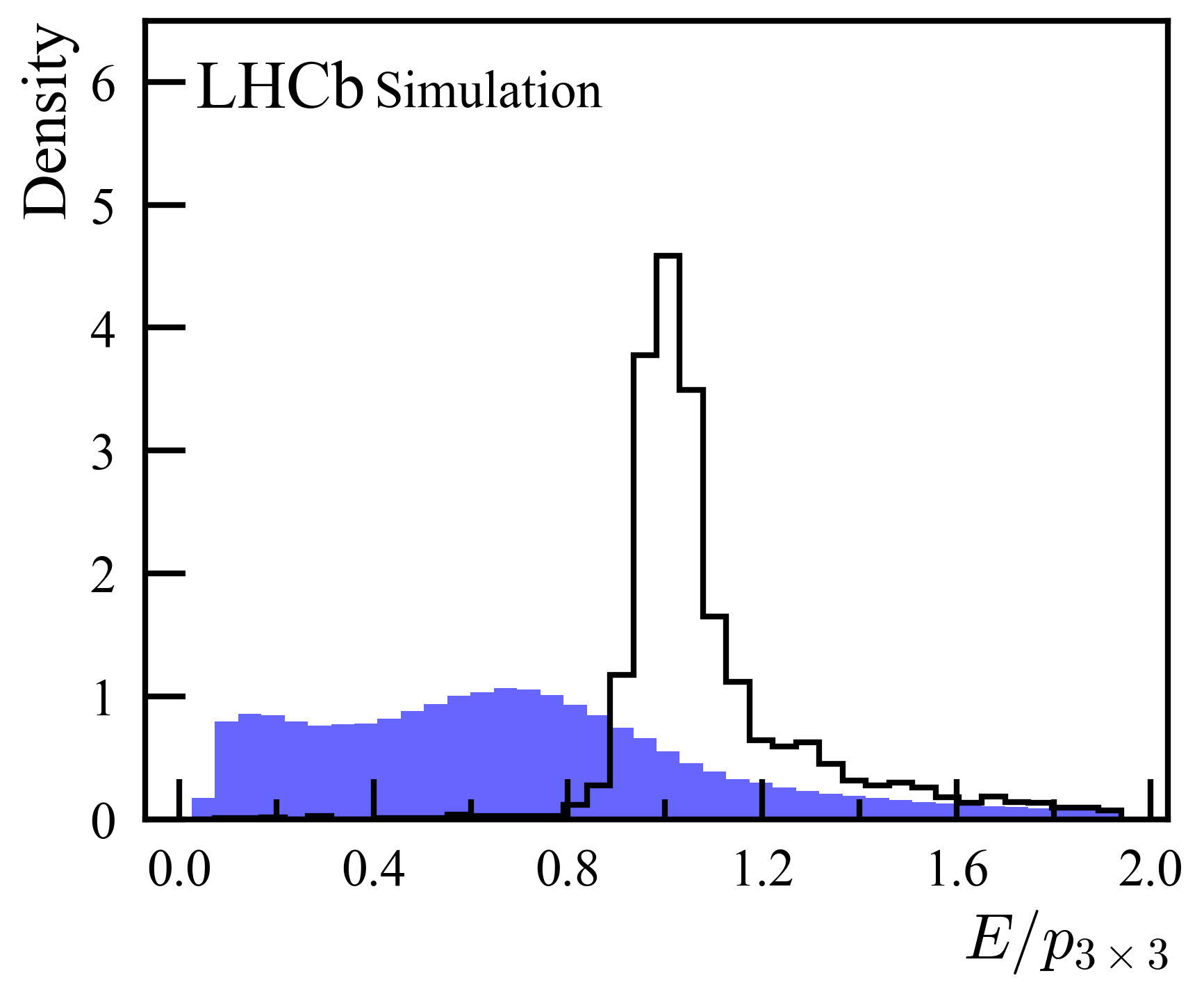}
\includegraphics[width=0.32\textwidth]{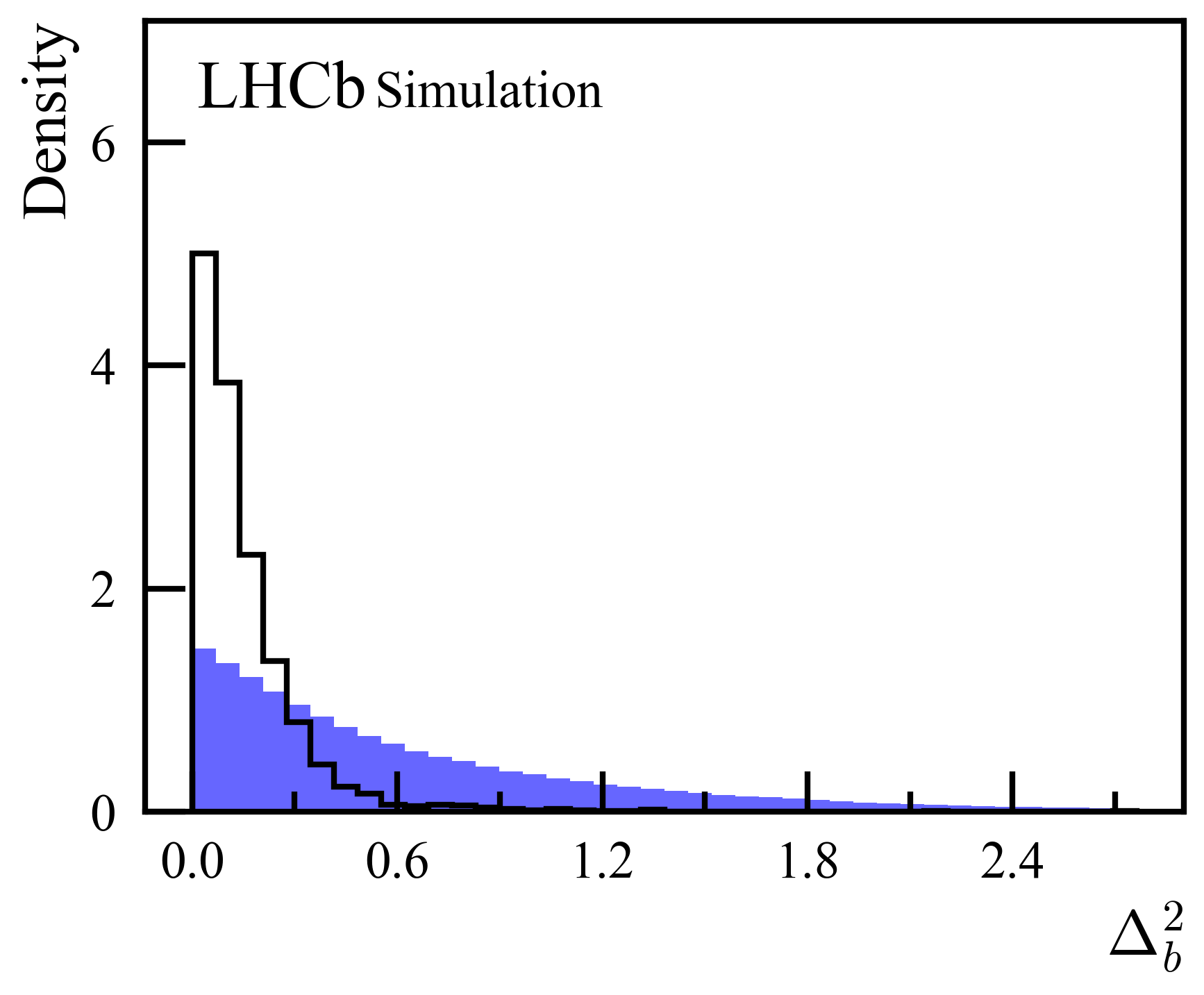}
\includegraphics[width=0.32\textwidth]{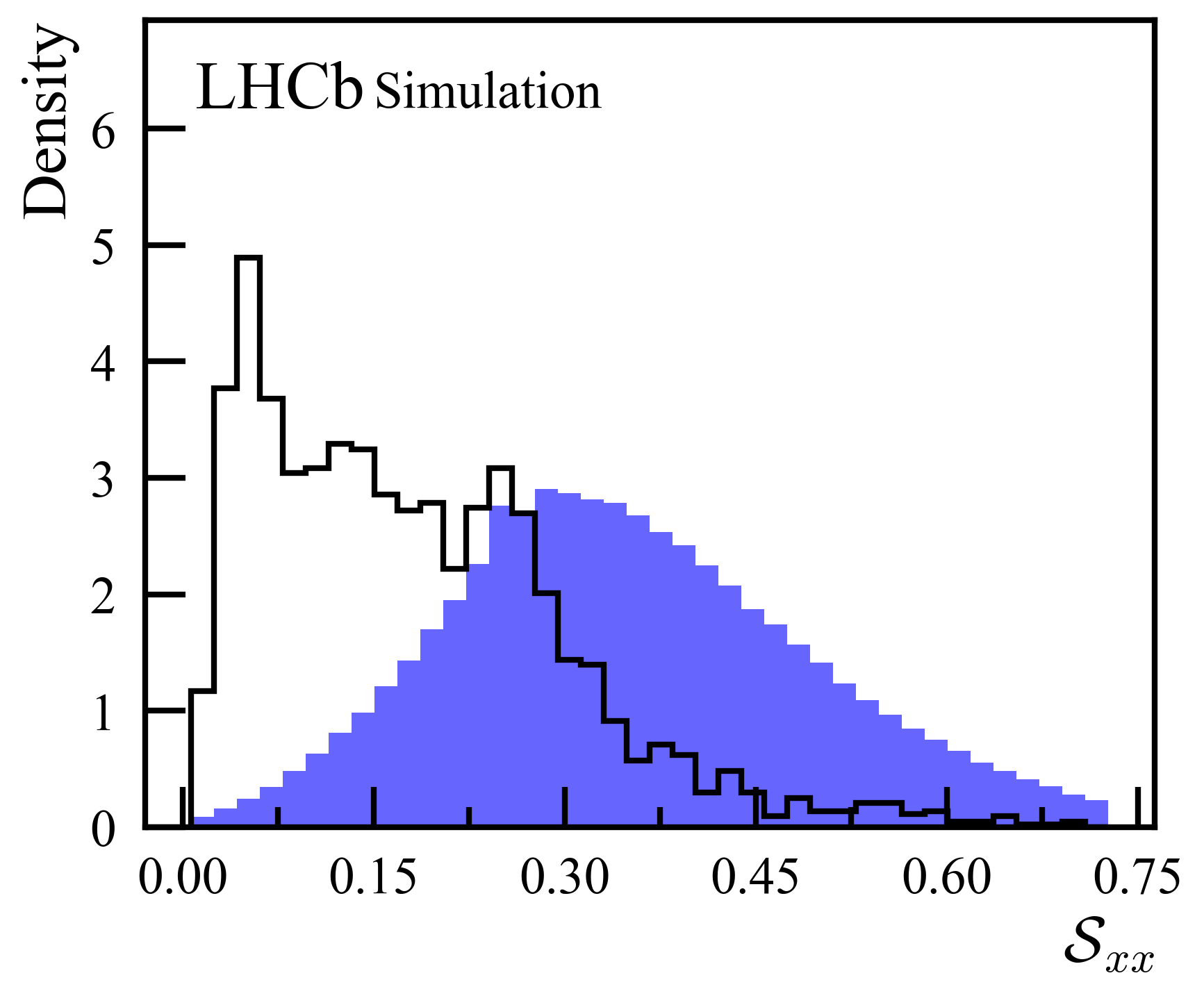}
\includegraphics[width=0.32\textwidth]{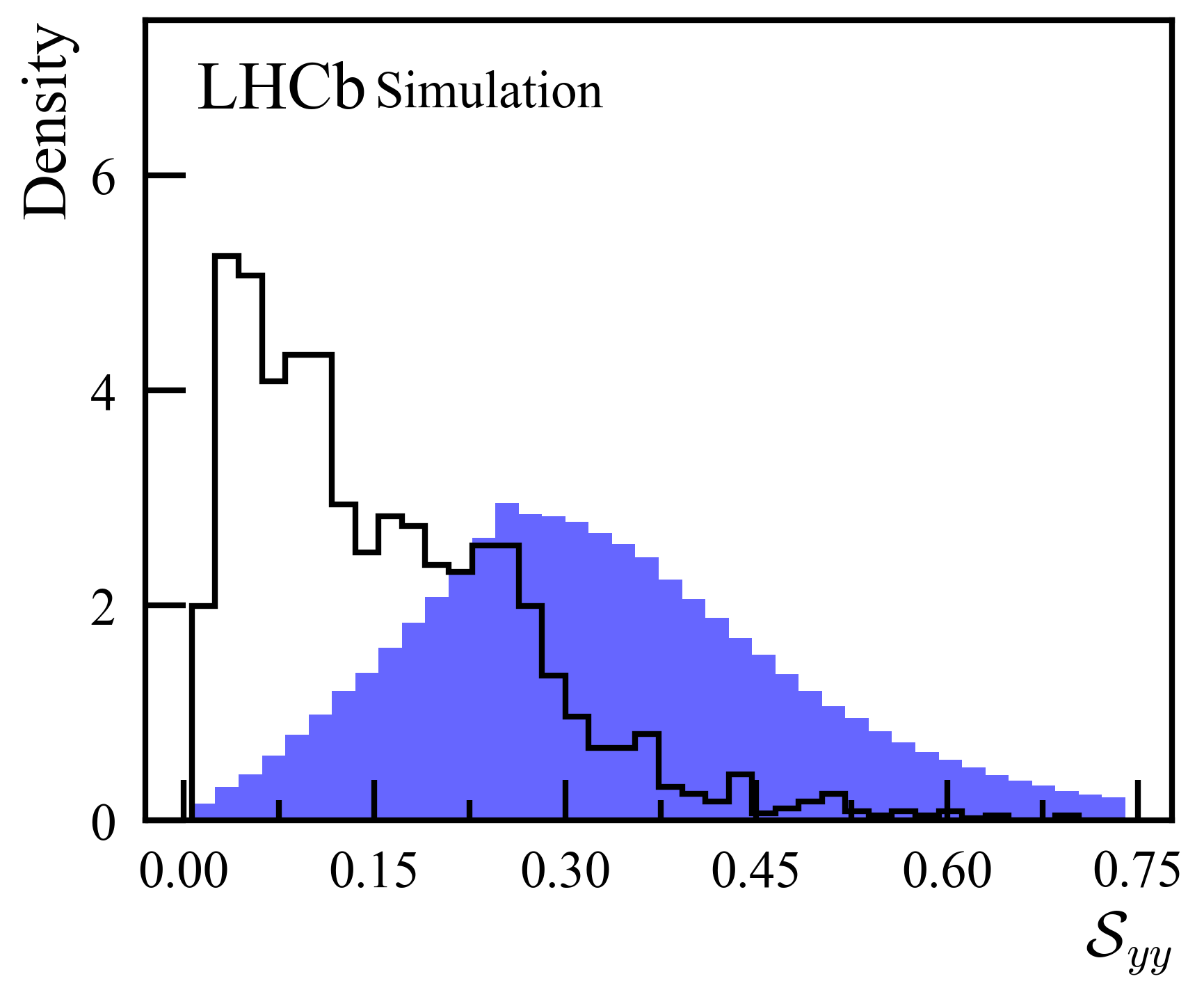}
\includegraphics[width=0.32\textwidth]{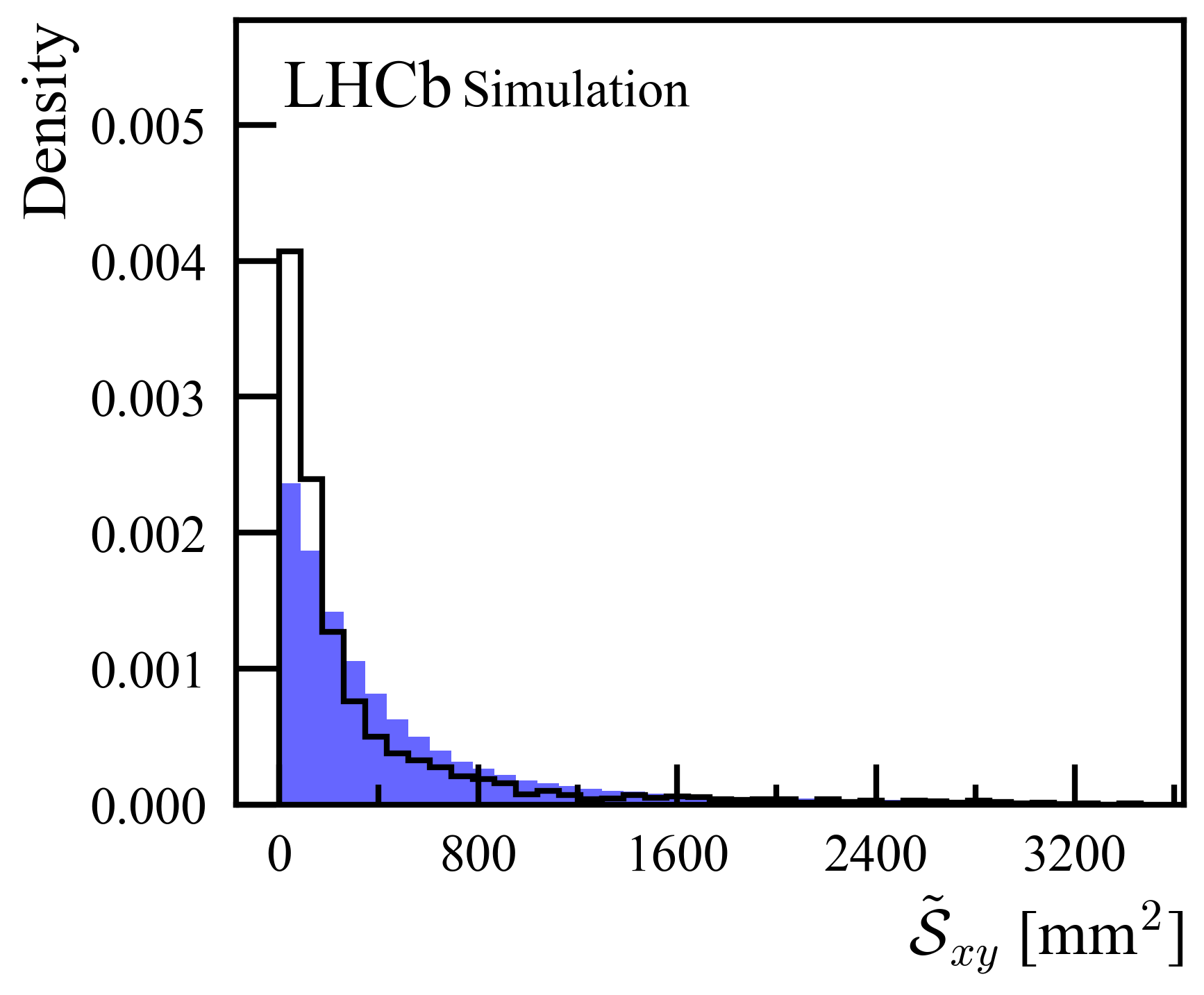}
\caption{\label{fig:e-features} Distributions of the six input features used by the electron-identification Lipschitz neural network, shown for simulated electrons (black) and pions (blue) in the momentum bin $p\in[6,10]\,\mathrm{GeV}$.}
\end{figure}

The Lipschitz-constrained neural network combines the ECAL-based observables described above to provide robust electron identification across a wide momentum range while remaining compatible with the throughput and memory constraints of real-time GPU execution.
Figure~\ref{fig:e-features} shows the distributions of the six input features used in the electron-identification Lipschitz-constrained neural network (prior to rescaling). 

\section{Performance}
\label{sec:perform}

The performance of the Lipschitz-constrained neural network PID algorithms is evaluated using simulated data and compared to the baseline cut-based trigger PID. 
The comparison is performed in terms of signal efficiency and background rejection across the relevant kinematic phase space. 
In all cases considered, the neural-network-based operators provide improved discrimination relative to the baseline methods.

\begin{figure}[t]
    \centering
    \includegraphics[width=0.32\textwidth]{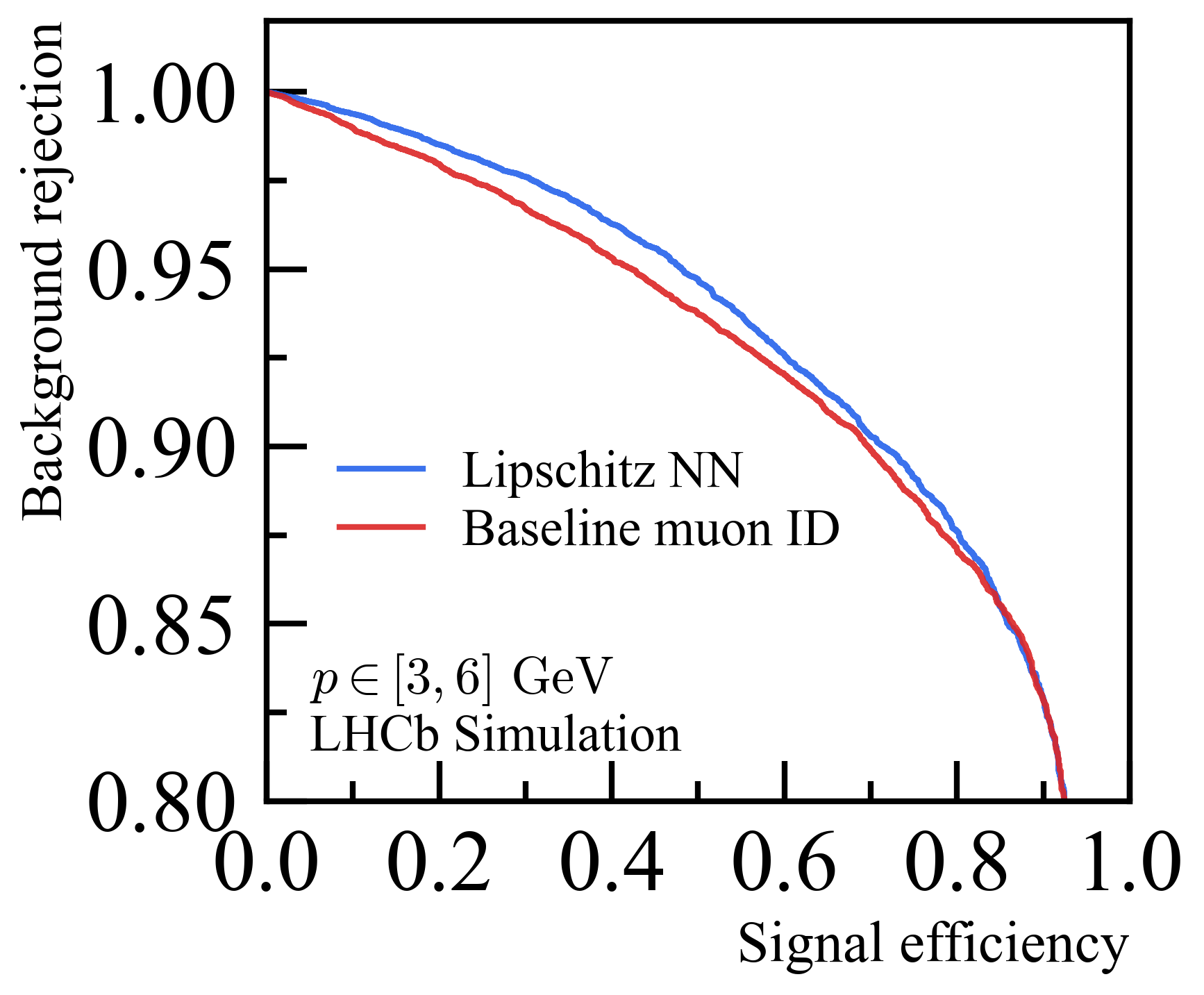}
    \includegraphics[width=0.32\textwidth]{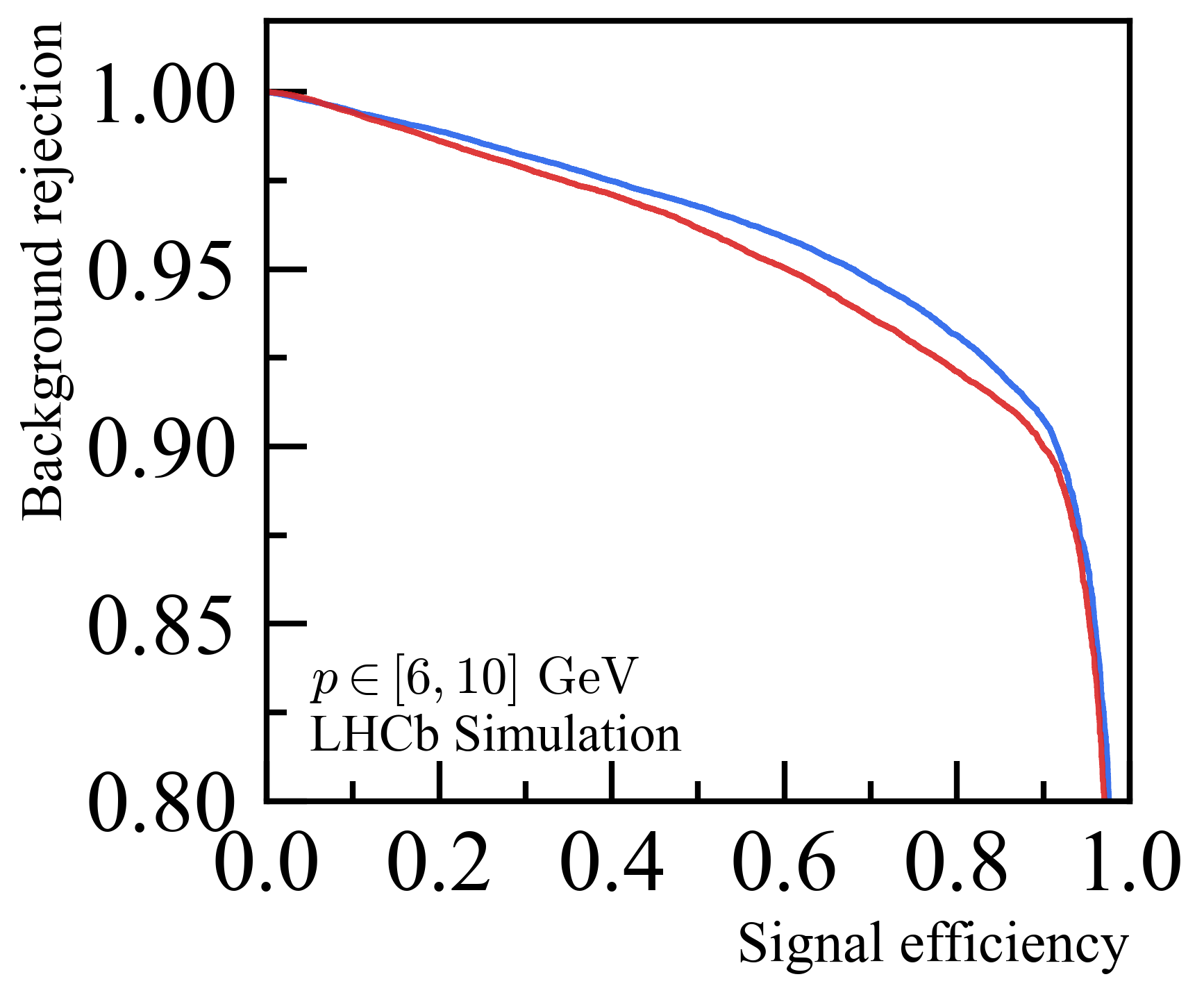}
    \includegraphics[width=0.32\textwidth]{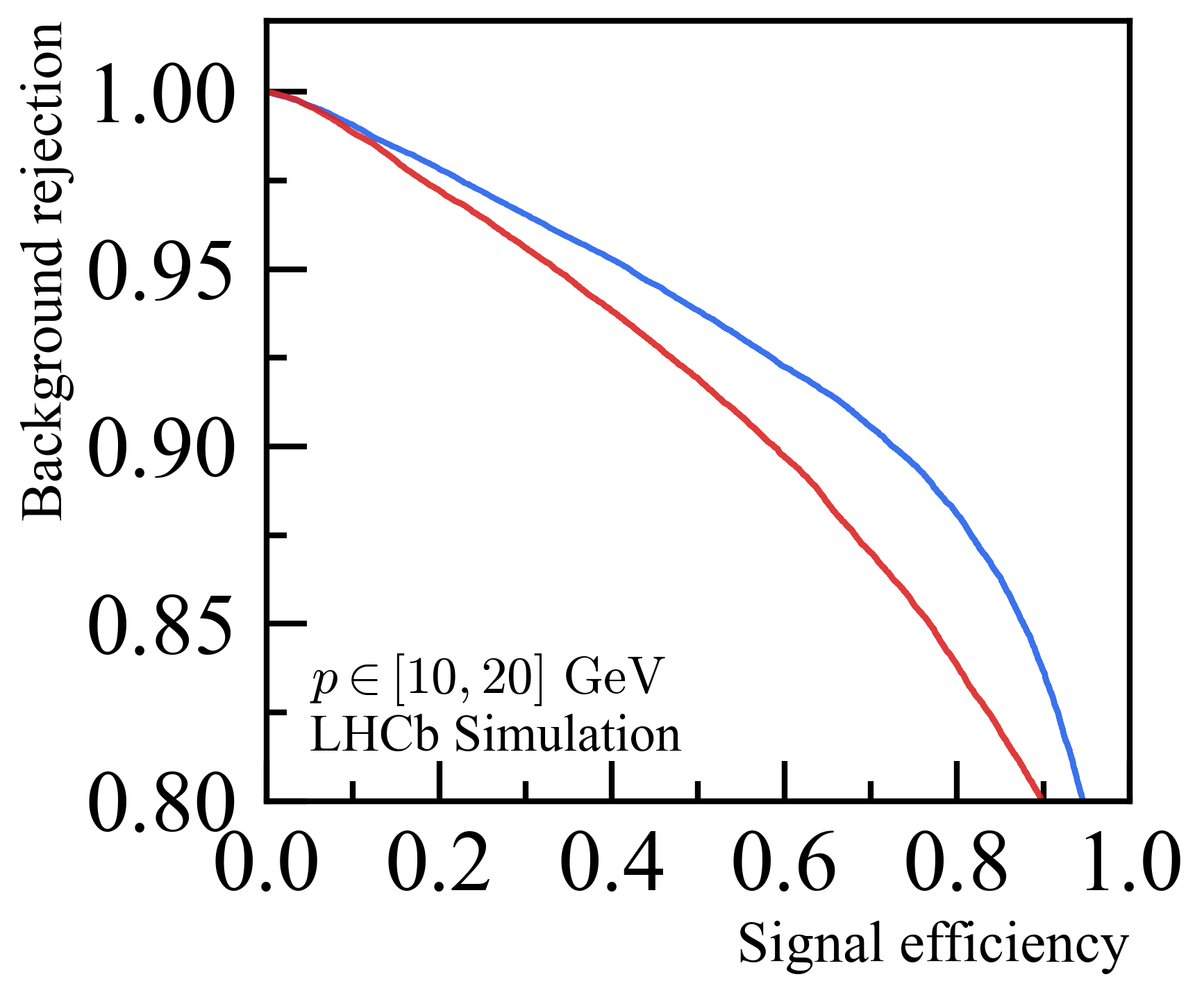}
    \includegraphics[width=0.32\textwidth]{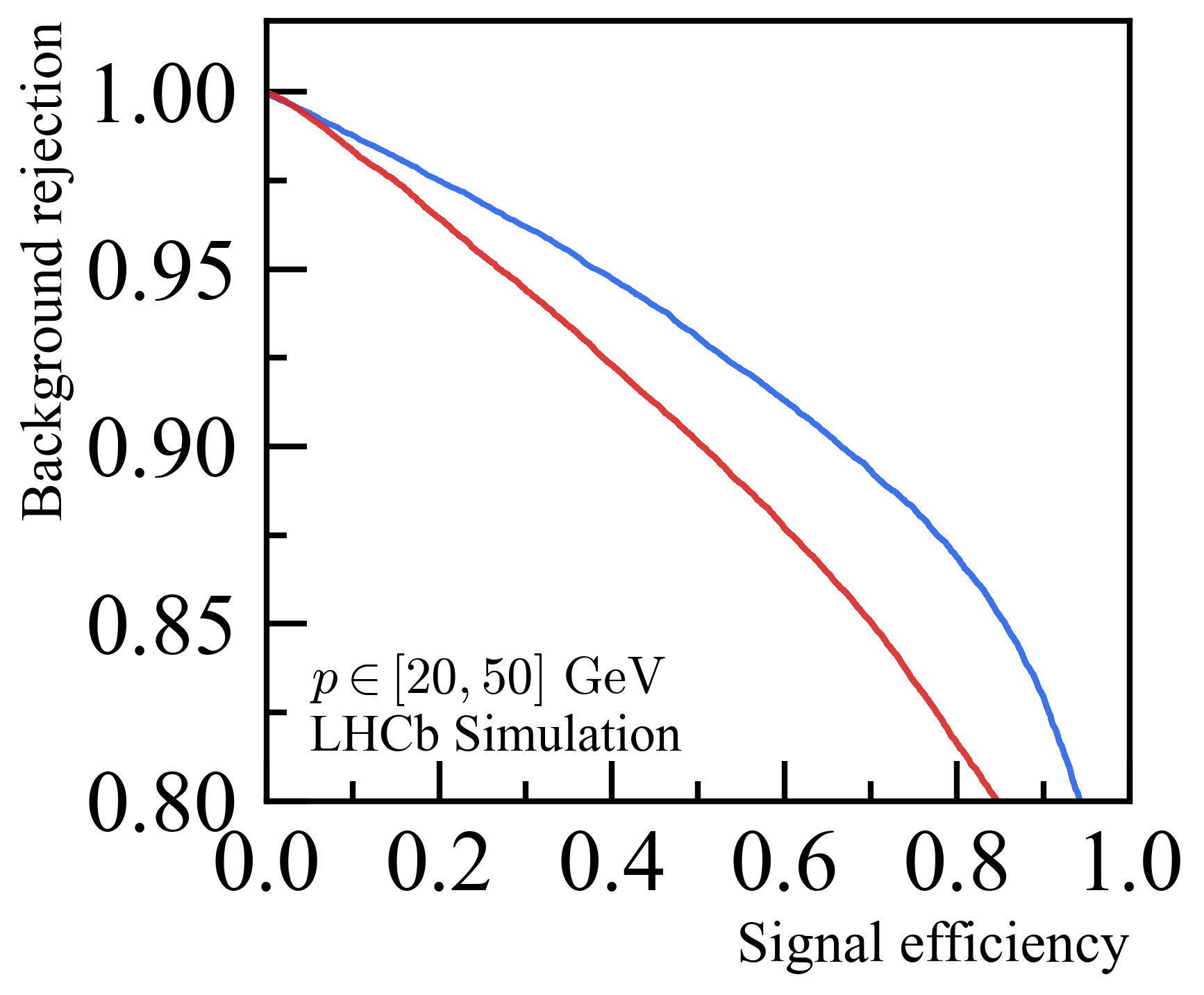}
    \includegraphics[width=0.32\textwidth]{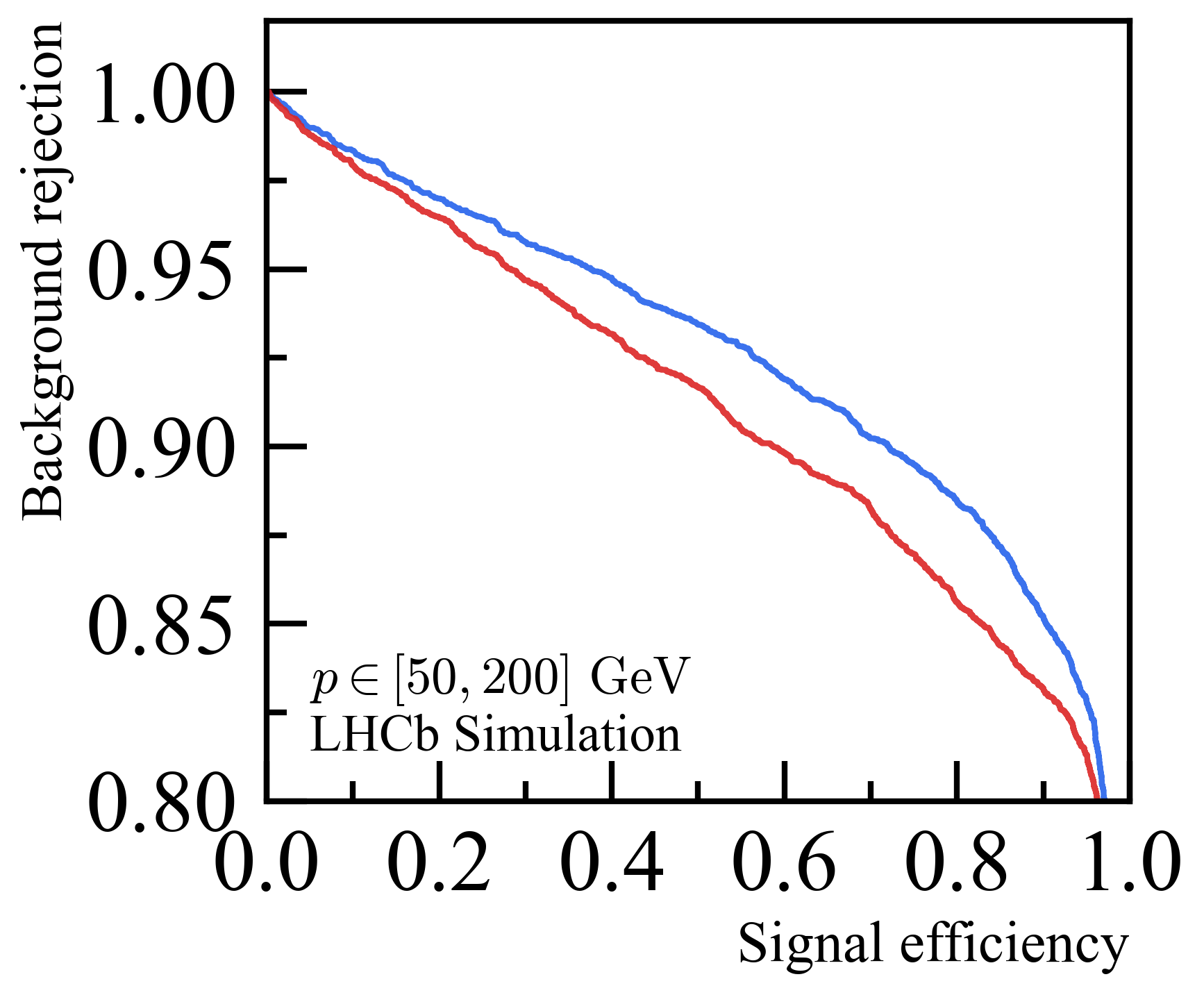}
    \caption{Background rejection versus signal efficiency for the Lipschitz-constrained neural network and the baseline muon ID~\cite{LHCbMuonID2020}, evaluated on simulated muons and pions in bins of momentum.}
    \label{fig:roc-muon-2d}
\end{figure}

For muon identification, the Lipschitz-constrained neural network achieves strong performance over the full momentum range relevant to the HLT1 trigger. 
Figure~\ref{fig:roc-muon-2d} shows the performance in bins of momentum $p$.
Substantial gains are achieved for moderate-to-high momenta, which includes most tracks that satisfy the kinematic-based HLT1 pre-selection criteria. 
This enhanced pion rejection is essential for maintaining manageable trigger output rates while preserving high efficiency for heavy-flavor and exotic decays containing muons in the final state.
For muons with lower momenta, the performance of the baseline and neural-network PID algorithms is found to be similar. 

\begin{figure}[t]
    \centering
    \includegraphics[width=0.32\textwidth]{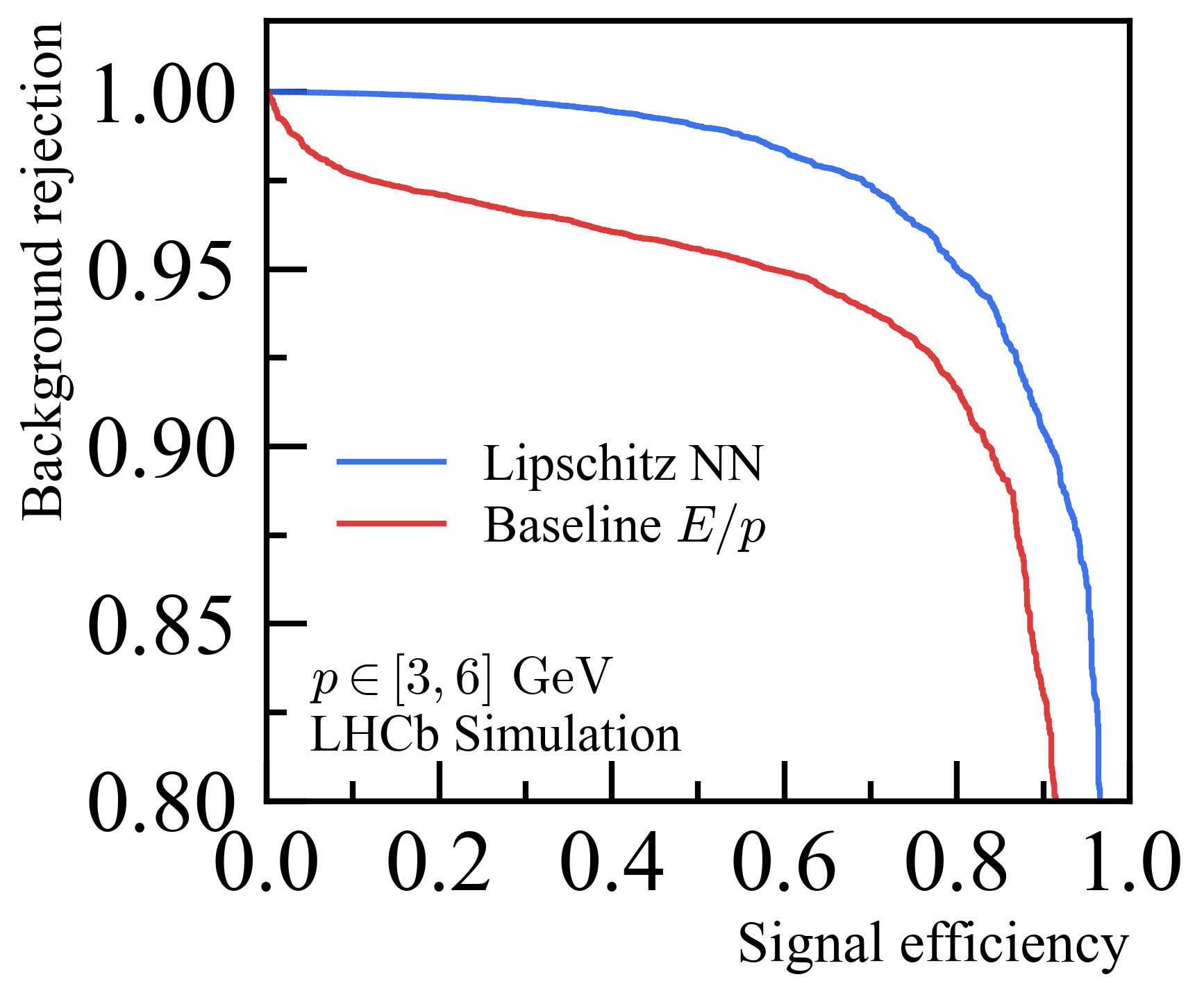}
    \includegraphics[width=0.32\textwidth]{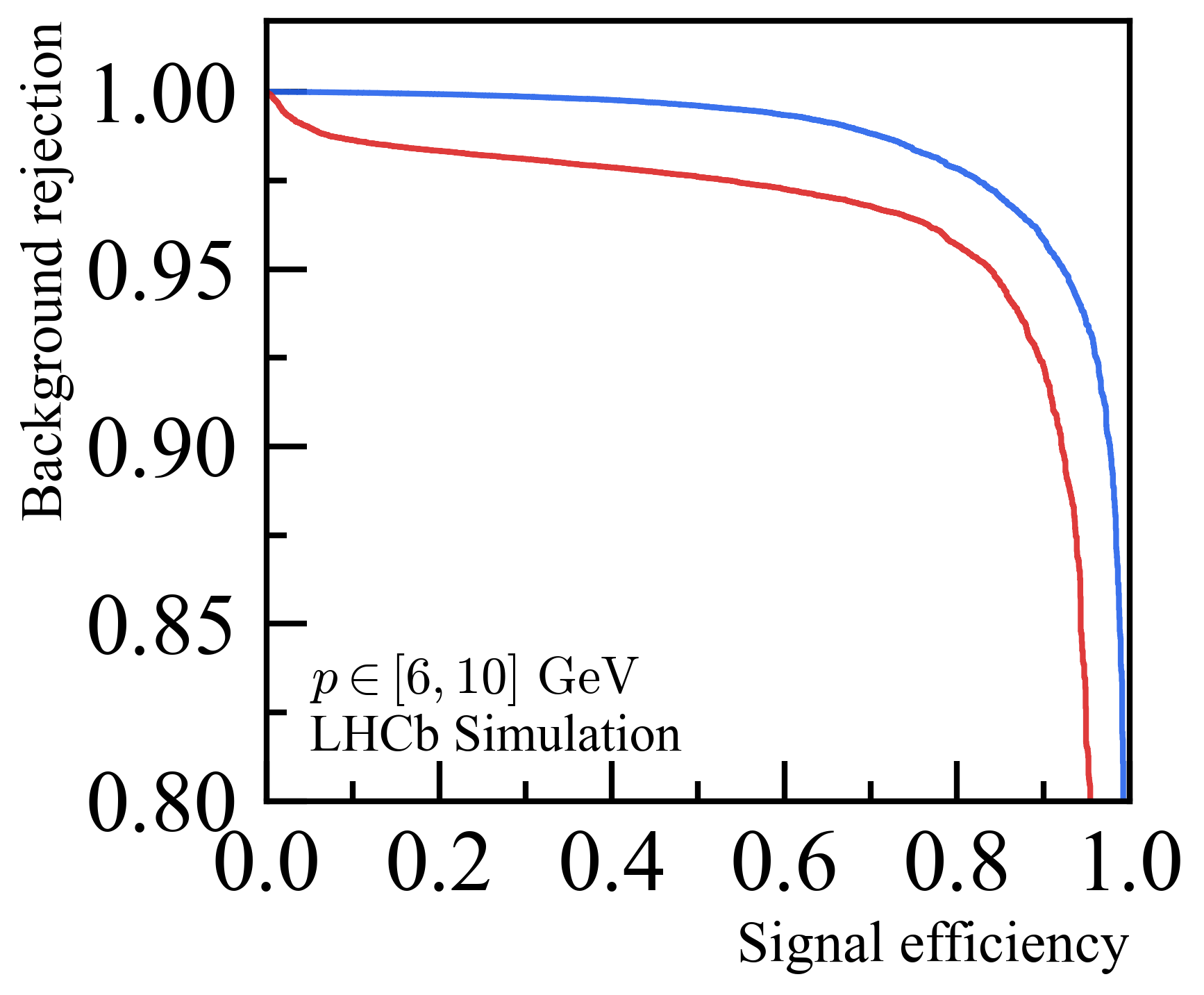}
    \includegraphics[width=0.32\textwidth]{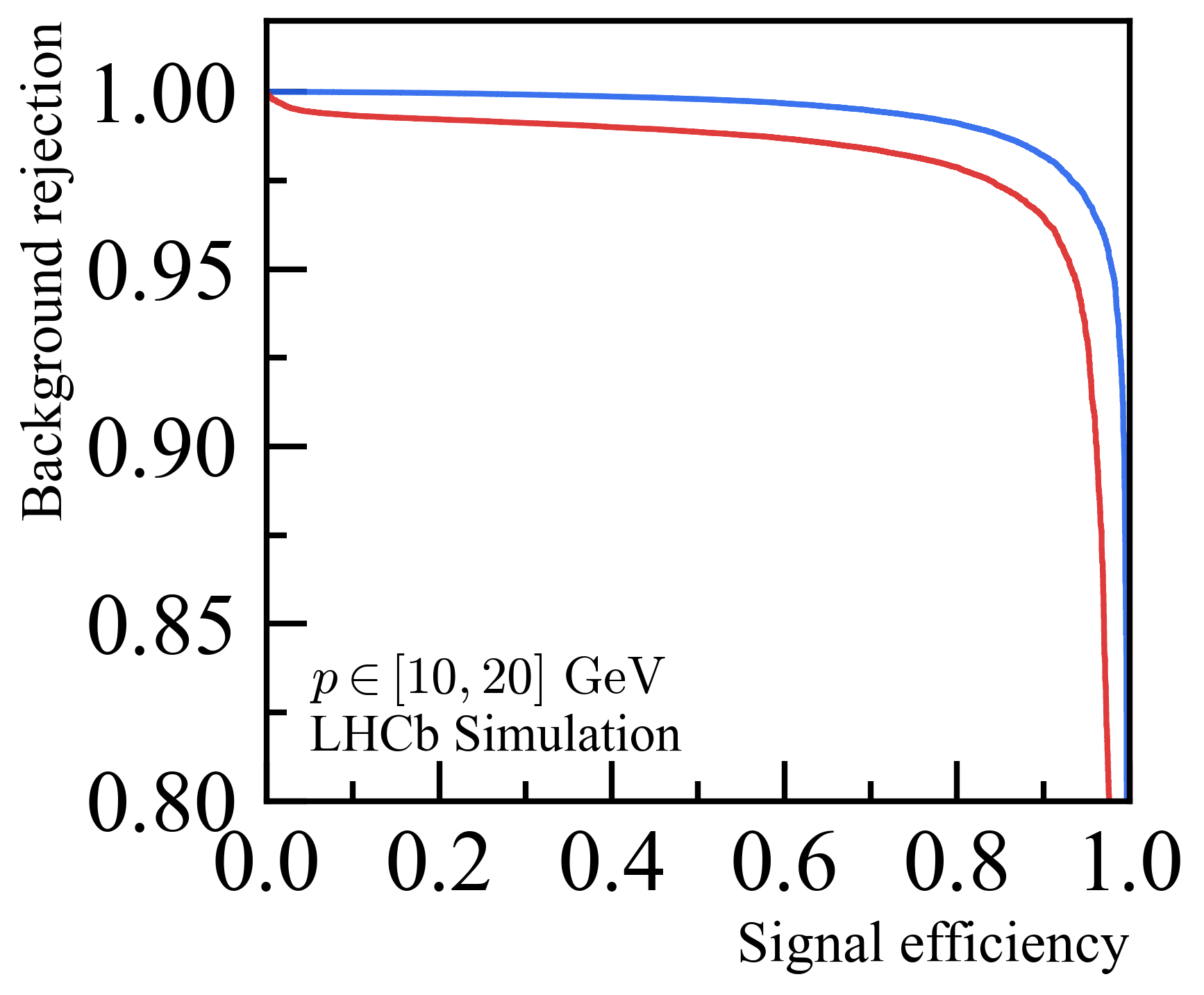}
    \includegraphics[width=0.32\textwidth]{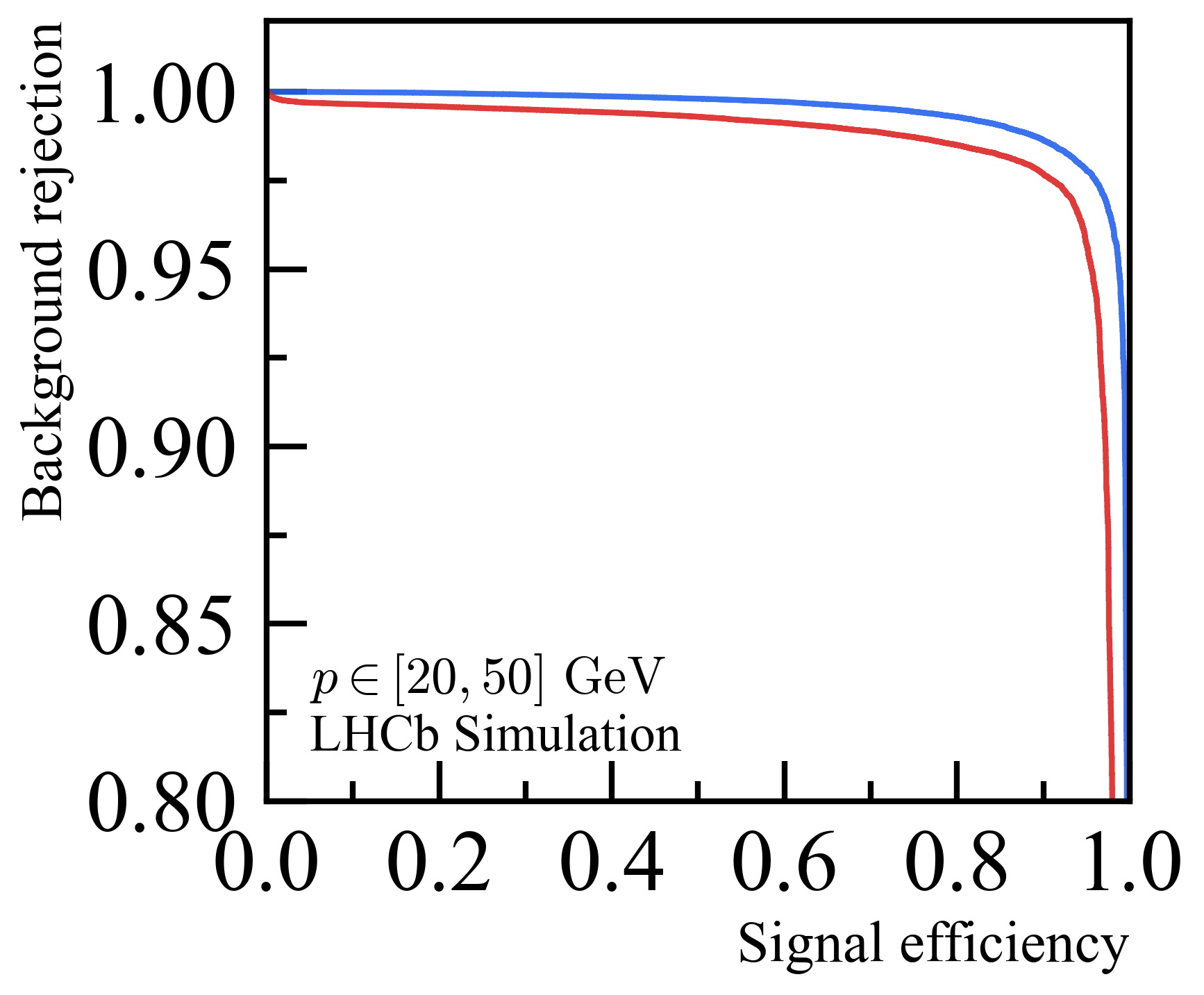}
    \includegraphics[width=0.32\textwidth]{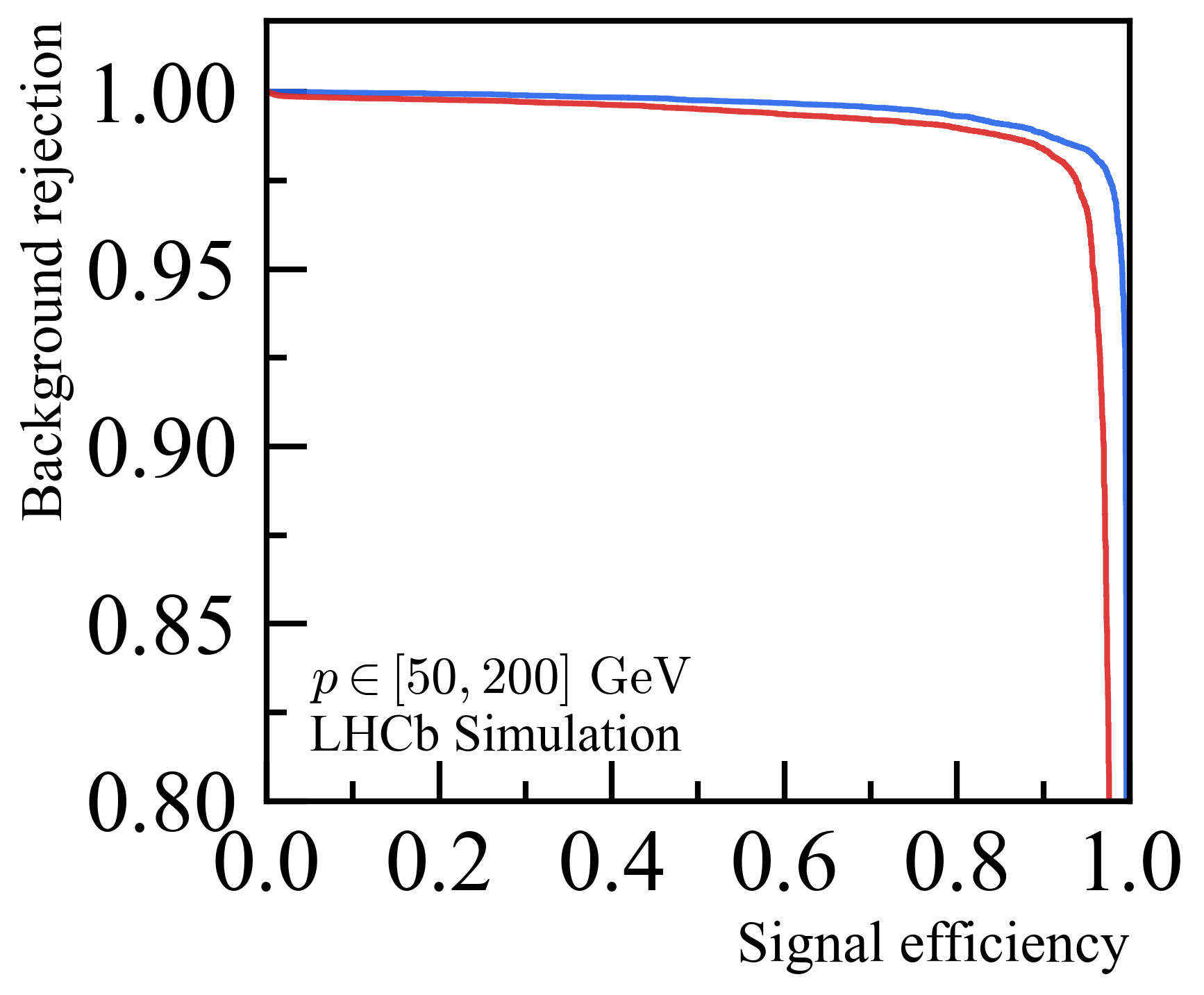}
    \caption{Background rejection versus signal efficiency for the Lipschitz-constrained neural network and the baseline $E/p$, evaluated on simulated electrons and pions in bins of momentum.}
    \label{fig:roc-electron-2d}
\end{figure}

For electron identification, the neural network provides robust discrimination against hadronic backgrounds across the full kinematic range studied. 
The improvement over the cut-based baseline is most pronounced at low momentum, where the ECAL energy resolution is worst and the discrimination provided by a single observable such as $E/p$ is weaker. 
The electron network shows a much weaker dependence on pseudorapidity and $p_\mathrm{T}$ with nearly uniform performance across the acceptance, and retains its improvement over the baseline in every kinematic bin studied. 
By combining information from track-cluster matching, shower-shape variables, and energy measurements, the neural network adapts naturally to the different physical regimes encountered across the electron momentum spectrum.
Figure~\ref{fig:roc-electron-2d} shows the performance in bins of momentum $p$.

\begin{figure}[t]
    \centering
    \includegraphics[width=0.6\textwidth]{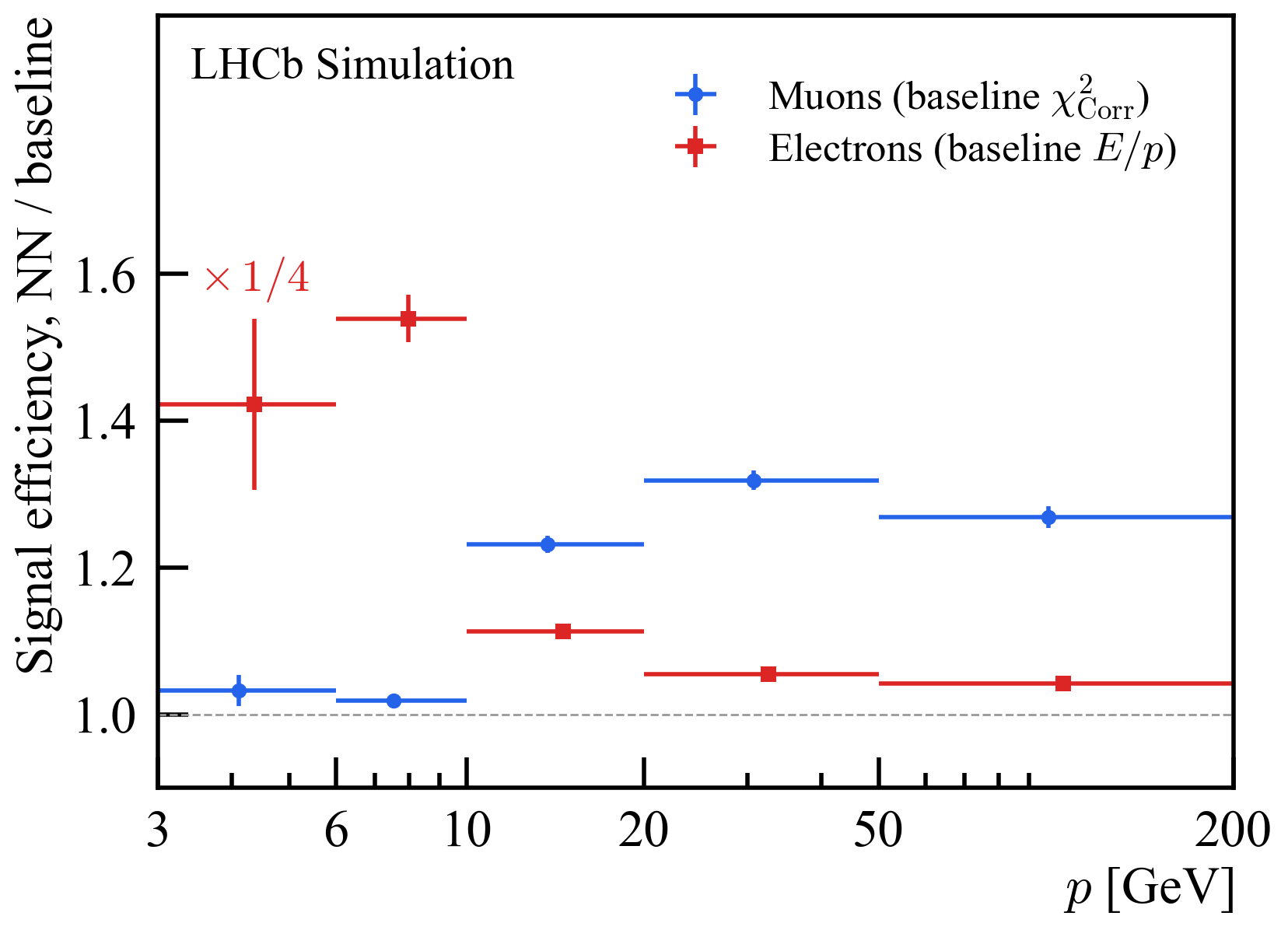}
    \caption{Ratio of the signal efficiency of the neural network to that of the baseline at a fixed background misidentification rate in each momentum bin: muons at $10\%$ pion misidentification (baseline $\chi^2_{\rm Corr}$~\cite{LHCbMuonID2020}) and electrons at $2.5\%$ hadron misidentification (baseline $E/p$). The vertical bars show the statistical uncertainties and the horizontal bars the bin widths. The lowest-momentum electron bin value is scaled down by a factor of 4 to make it visible.}
    \label{fig:eff-vs-p}
\end{figure}

Figure~\ref{fig:eff-vs-p} summarizes the efficiency gain of the neural network with respect to the baseline selections as a function of momentum at a fixed background misidentification rate, the natural figure of merit for a trigger, whose output rate is largely determined by the background rate. The ratio of the signal efficiency of the neural network to that of the baseline is shown, at a misidentification rate of $10\%$ for pions in muon identification and $2.5\%$ for hadrons in electron identification. These rates lie within the range of the working points of the HLT1 selections that use the networks. The muon network is up to $1.3$ times more efficient than the baseline, with the largest gains above $10\,$GeV. The electron network is $1.04$ to $1.5$ times more efficient above $6\,$GeV, and almost six times more efficient below $6\,$GeV, where $E/p$ retains only $12\%$ of the electrons, against $69\%$ for the network.

An important feature of both PID algorithms is their compact architecture. 
The neural networks are deliberately kept shallow, with four to five fully connected layers, to ensure compatibility with the throughput and memory constraints of the HLT1 GPU trigger. 
In the throughput test of the standard HLT1 sequence on an NVIDIA RTX A5000 GPU, which processes $78\times10^3$ events per second, corresponding to $13\,\mu$s per event, the muon- and electron-identification network kernels take $0.7\%$ and $0.6\%$ of the GPU time, about $0.09$ and $0.08\,\mu$s per event. For comparison, the IsMuon algorithm, which the muon network complements, takes $0.31\,\mu$s per event, and the computation of $\chi^2_{\rm Corr}$ and of the track--cluster matching variables, including $E/p$, takes $0.03$ and $0.08\,\mu$s per event, respectively. The networks themselves require only $\mathcal{O}(500)$ parameters (about 2~kB in single precision), and add 24–28 bytes of data per track to the batched track buffers. This constitutes a negligible addition to both the GPU memory pool and the per-track data volume already carried through HLT1 reconstruction.
Despite their small size, the networks achieve excellent discrimination power by exploiting correlated information from multiple input features. 
The explicit Lipschitz constraint further enforces controlled model behavior, which is particularly relevant in a real-time trigger context where inputs often differ slightly from those encountered during training.
Finally, although these neural network algorithms were trained on simulated data, they have been deployed and validated in the HLT1 trigger and have been running stably in production.  
These results demonstrate that Lipschitz-constrained neural networks provide a robust and efficient improvement over existing trigger-level PID methods, while remaining fully compatible with the real-time requirements of the LHCb Run~3 trigger.

\section{Balance between performance and robustness}
\label{sec:robust}

To quantify the robustness provided by the Lipschitz constraint, the deployed networks are compared with otherwise identical networks whose Lipschitz bounds are so large that these networks are effectively unconstrained. The simulated detector response is distorted and the network inputs are recomputed. For muons, stations M2 and M3 are displaced along $x$ by $+\delta$ and $-\delta$, with $\delta$ up to $20\,$mm. Displacements of this size are not expected during data taking~\cite{muon-performancerun3}: 
therefore, this scan is a stress test well beyond the expected misalignments. 
For electrons, the transverse width of the ECAL showers is scaled by a common factor. The three second moments, which grow with the square of the shower width, are scaled by the square of this factor. The two $E/p$ values are reduced by the energy that a wider shower deposits outside the crossed cells and outside the $3\times3$ region, respectively; this loss is computed separately for each ECAL region, with a simple model of the transverse shower profile.

\begin{figure}[t]
    \centering
    \includegraphics[width=0.49\textwidth]{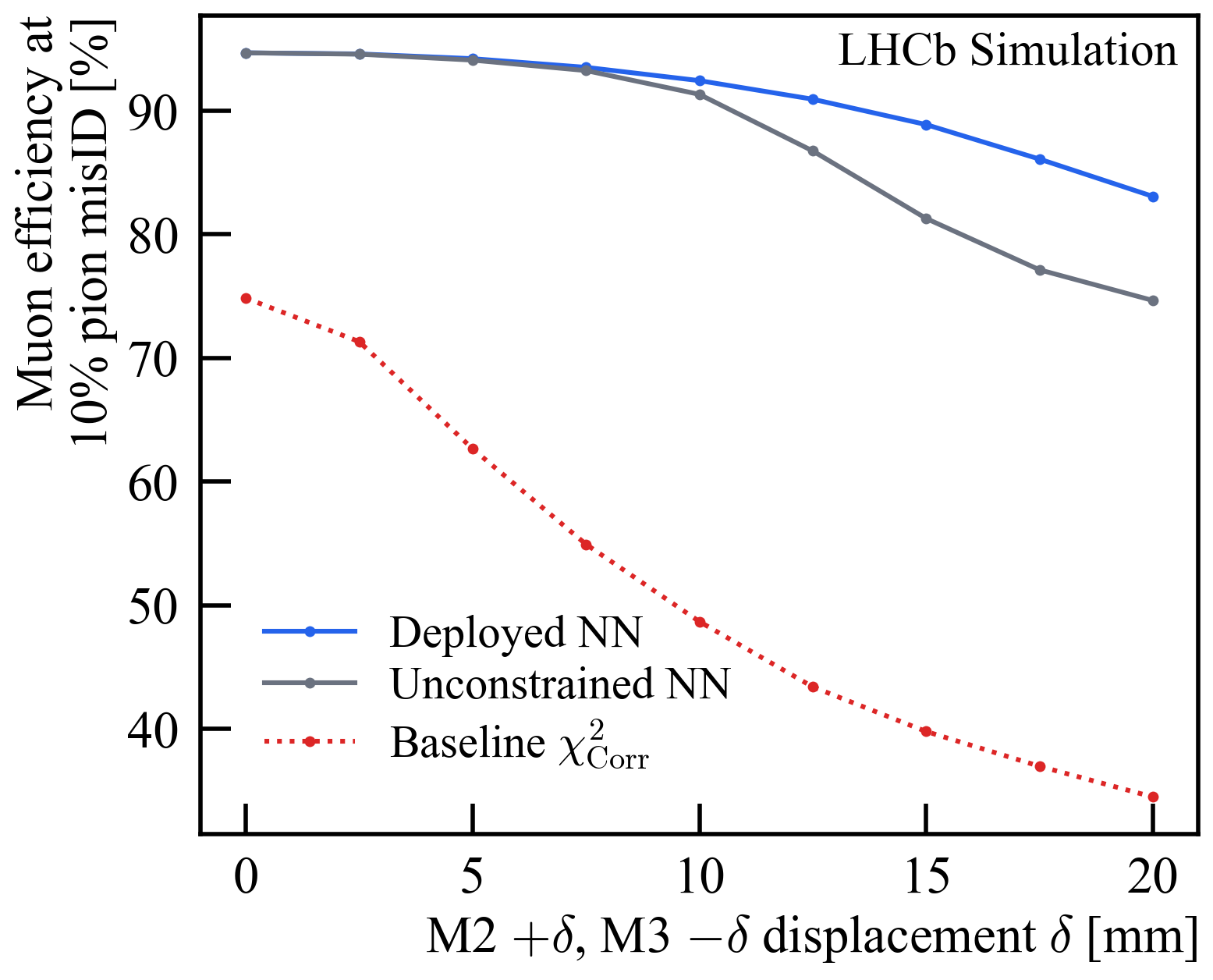}
    \hfill
    \includegraphics[width=0.49\textwidth]{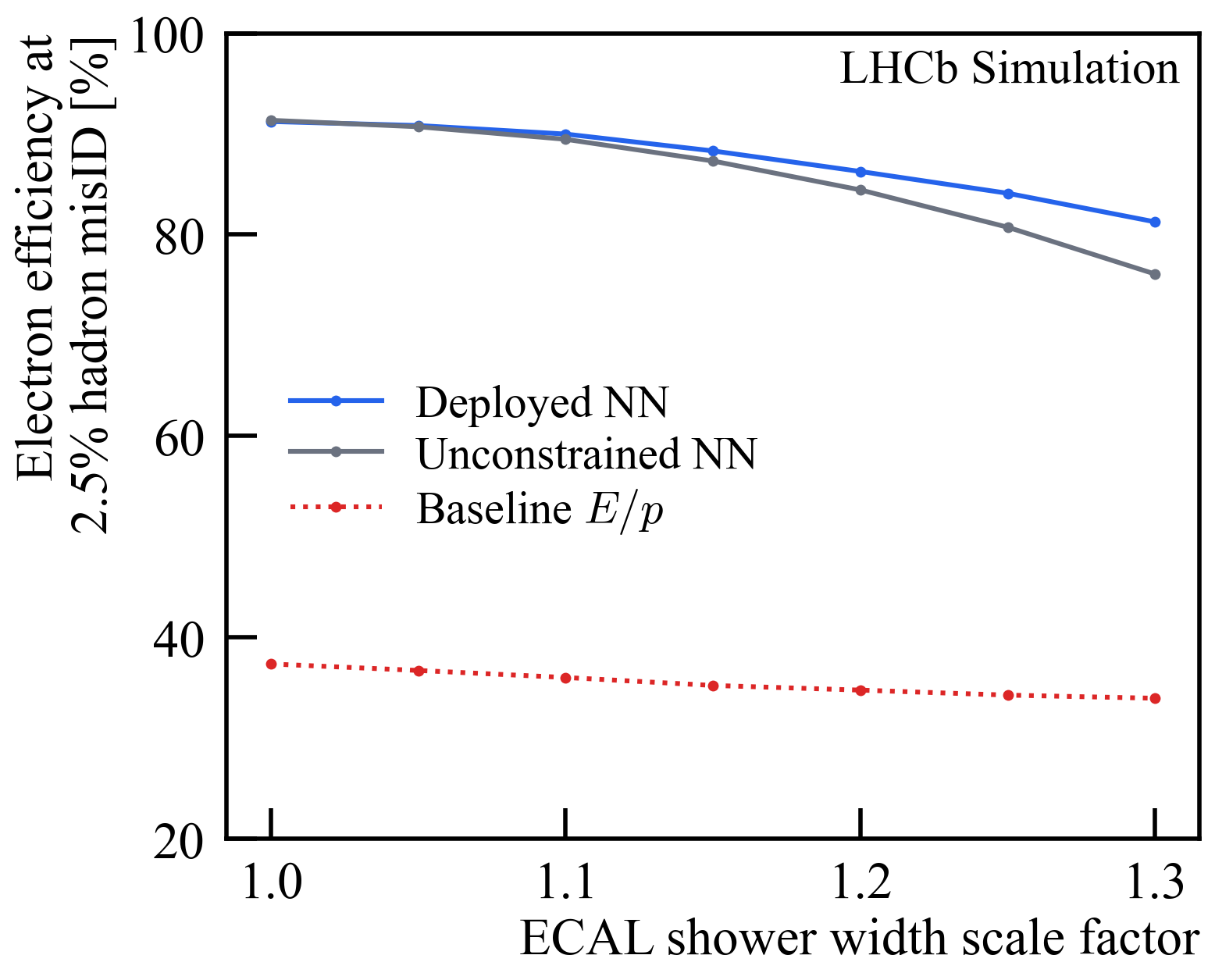}
    \caption{Signal efficiency at a fixed background misidentification rate for a distorted detector response, for the deployed network, an otherwise identical unconstrained network, and the baseline. Left: muon efficiency at $10\%$ pion misidentification for stations M2 and M3 displaced along $x$ by $+\delta$ and $-\delta$; baseline $\chi^2_{\rm Corr}$~\cite{LHCbMuonID2020}. Right: electron efficiency at $2.5\%$ hadron misidentification as a function of a scale factor applied to the width of the ECAL showers, which changes the second moments and both $E/p$ values; baseline $E/p$.} %
    \label{fig:robustness}
\end{figure}

Figure~\ref{fig:robustness} shows the signal efficiency at a fixed misidentification rate as a function of the distortions. In nominal conditions the deployed and unconstrained networks perform identically, and small distortions barely affect either, while the efficiency of $\chi^2_{\rm Corr}$ drops from $75\%$ to $49\%$ for a $1\,$cm displacement. For larger distortions the deployed networks retain a higher efficiency: $83\%$ against $75\%$ for muons at $\delta=20\,$mm, and $81\%$ against $76\%$ for electrons with showers $30\%$ wider, for which the efficiency of $E/p$ drops from $37\%$ to $34\%$. A tighter bound would add robustness against such large distortions, at a cost in discrimination under nominal conditions, thus the deployed networks were not constrained further. Tighter bounds may become worthwhile at the High-Luminosity LHC, where the detector response is expected to degrade significantly during data taking.

\section{Summary and discussion}

In this article, we have presented the development and application of Lipschitz-constrained neural networks for real-time muon and electron identification in the LHCb Run~3 trigger. The LHCb Run 3 detector now operates with a fully software-based trigger that processes the complete detector readout at an input rate of approximately 30~MHz, placing stringent requirements on the efficiency, throughput, and robustness of trigger-level particle-identification algorithms.
Neural network architectures have been developed for both muon and electron identification, with model complexity explicitly constrained to ensure compatibility with the throughput and memory footprint requirements of the HLT1 GPU trigger. 
The use of Lipschitz constraints provides controlled model behavior and bounded sensitivity to variations in the input features, which is particularly advantageous in a real-time environment where detector conditions and event characteristics can evolve over time.

For muon identification, the neural network combines information from track-muon matching observables and track-slope comparisons between detector subsystems to discriminate genuine muons from hadrons that penetrate the absorbers or decay in flight. 
The resulting classifier achieves strong pion rejection while maintaining high muon efficiency across the kinematic range relevant to HLT1 operation. 
For electron identification, the neural network exploits calorimeter-based observables, including energy-over-momentum ratios, track-cluster matching, and shower-shape variables, to separate electrons from hadronic backgrounds. 
The improvement over baseline cut-based methods is particularly pronounced in kinematic regions where individual observables provide limited discrimination.

The performance presented here is based on simulated data and demonstrates a consistent improvement over the existing trigger-level PID. 
Though not presented here, the neural network algorithms have been deployed and operationally validated within the HLT1 trigger and have been running stably in production. 
The successful deployment of these algorithms demonstrates that Lipschitz-constrained neural networks can meet the practical requirements of real-time trigger operation in a high-rate environment.
Our approach illustrates the potential of carefully constrained machine-learning models for use in critical online systems, where both performance and reliability are essential. 
Beyond their immediate application to lepton identification at LHCb, the techniques developed in this work may be relevant to other real-time decision-making problems in high-energy physics experiments or other domains.

\acknowledgments
We thank the LHCb Simulation, Computing, Online, and Real-Time Analysis (RTA) projects for their support in producing the samples, providing computing resources, and supporting and reviewing this publication. 
We thank Carla Marin Benito and Lorenzo Paolucci for useful suggestions and feedback. 
The MIT group was supported by NSF grants PHY-2019786 (The NSF AI Institute for Artificial Intelligence and Fundamental Interactions, http://iaifi.org/),  PHY-2209181, PHY-2411593, and PHY-2512443. 
MS acknowledges support from the Istituto Nazionale di Fisica Nucleare (INFN), Laboratori Nazionali di Frascati. MF acknowledges support from the INFN, Sezione di Bologna.

\bibliographystyle{JHEP}
\bibliography{biblio.bib}

\end{document}